# The CluMPR galaxy cluster-finding algorithm and DESI legacy survey galaxy cluster catalogue

M. J. Yantovski-Barth,[1,2,3,4]★ Jeffrey A. Newman,[1] Biprateep Dey,[1] Brett H. Andrews,[1] Michael Eracleous,[5] Jesse Golden-Marx[6] and Rongpu Zhou[7]

[1]*Department of Physics and Astronomy and Pittsburgh Particle Physics, Astrophysics and Cosmology Center (PITT PACC), University of Pittsburgh, Pittsburgh, PA 15260, USA*
[2]*Department of Physics, Université de Montréal, Complexe des sciences, 1375 Avenue Thérèse-Lavoie-Roux, Montréal (Québec) H2V 0B3, Canada*
[3]*Ciela Institute, Université de Montréal, Complexe des sciences, 1375 Avenue Thérèse-Lavoie-Roux, Montréal (Québec) H2V 0B3, Canada*
[4]*Mila, U6666 St-Urbain Street, #200, Montréal (Québec) H2S 3H1, Canada*
[5]*Department of Astronomy and Astrophysics and Institute for Gravitation and the Cosmos, Pennsylvania State University, 525 Davey Laboratory, University Park, PA 16802, USA*
[6]*Department of Astronomy, School of Physics and Astronomy, Shanghai Jiao Tong University, Shanghai 200240, China*
[7]*Lawrence Berkeley National Laboratory, 1 Cyclotron Road, Berkeley, CA 94720, USA*

**ABSTRACT**
Galaxy clusters enable unique opportunities to study cosmology, dark matter, galaxy evolution, and strongly lensed transients. We here present a new cluster-finding algorithm, CluMPR (Clusters from Masses and Photometric Redshifts), that exploits photometric redshifts (photo-$z$'s) as well as photometric stellar mass measurements. CluMPR uses a 2D binary search tree to search for overdensities of massive galaxies with similar redshifts on the sky and then probabilistically assigns cluster membership by accounting for photo-$z$ uncertainties. We leverage the deep DESI Legacy Survey $grzW1W2$ imaging over one-third of the sky to create a catalogue of ∼ 300 000 galaxy cluster candidates out to $z = 1$, including tabulations of member galaxies and estimates of each cluster's total stellar mass. Compared to other methods, CluMPR is particularly effective at identifying clusters at the high end of the redshift range considered ($z = 0.75–1$), with minimal contamination from low-mass groups. These characteristics make it ideal for identifying strongly lensed high-redshift supernovae and quasars that are powerful probes of cosmology, dark matter, and stellar astrophysics. As an example application of this cluster catalogue, we present a catalogue of candidate wide-angle strongly lensed quasars in Appendix C. The nine best candidates identified from this sample include two known lensed quasar systems and a possible changing-look lensed QSO with SDSS spectroscopy. All code and catalogues produced in this work are publicly available (see Data Availability).

## 1 INTRODUCTION

Galaxy clusters are the most massive gravitationally bound structures in the Universe, and therefore they are of great use in cosmology and astrophysics. One motivation for the creation of catalogues of clusters is to locate likely areas for observing strongly lensed time-variable phenomena such as gravitationally lensed supernovae and quasars (Meyers et al. 2009). The observation of the time delays between the lensed images of variable phenomena, coupled with proper modelling of the lensing system, can provide an independent constraint on the Hubble–Lemaître parameter (Refsdal 1964). In 2015, multiple images of a supernova lensed by a galaxy cluster were observed (Kelly et al. 2015). Recently, the intermediate Palomar Transient Factory and its successor, the Zwicky Transient Facility (ZTF), have discovered several additional gravitationally lensed supernovae (Goobar et al. 2017; Goobar et al. 2023). Meanwhile, programmes to observe gravitationally lensed quasars are ongoing (Wong et al. 2020; Napier et al. 2023). Further observations of gravitationally lensed transients will lead to improved constraints on the nature of dark matter and the expansion of the Universe.

Galaxy cluster catalogues have many other applications. As one example, they can be used to study the impact of clusters on the evolution of galaxies in their vicinity (Dressler 1980). Gravitational magnification of background galaxies by galaxy clusters enables distant galaxies to be detected and studied at high resolution (e.g. Bayliss et al. 2014; Livermore et al. 2015; Johnson et al. 2017; Cornachione et al. 2018; Rivera-Thorsen et al. 2019; Ivison et al. 2020; Khullar et al. 2021; Bezanson et al. 2022). Furthermore, they can enable the detection and analysis of merging galaxy clusters, which are useful for constraining the dark matter scattering cross-section (Wittman et al. 2023). Additionally, measurements of the apparent abundance of clusters can be a sensitive probe of cosmology (Haiman, Mohr & Holder 2001; Newman et al. 2002).

★ E-mail: michael.barth@umontreal.ca

A purely photometric galaxy cluster-finding algorithm can take advantage of the large number of distant sources which can be observed using imaging. Conveniently, estimates of redshift and stellar mass can be obtained from the same photometric data. An algorithm which is optimally adapted for photometric observations will be essential for the upcoming generation of optical and infrared surveys at facilities such as Rubin Observatory (LSST Science Collaboration 2009), Euclid Observatory (Euclid Collaboration 2022), and the Nancy Grace Roman Telescope (Spergel et al. 2015), all of which will generate immense amounts of photometric data spanning many passbands.

The redMaPPer algorithm (Rykoff et al. 2014) is an example of a well-tested photometric galaxy cluster-finding algorithm. The redMaPPer algorithm relies on a red-sequence model to detect candidate galaxy clusters. Currently redMaPPer has been applied to data from the Sloan Digital Sky Survey Data Release 8 (SDSS DR8) (Aihara et al. 2011), Dark Energy Survey Science Verification (DES SV) (Dark Energy Survey Collaboration 2016), and Dark Energy Survey Year 1 (DES Y1) (Drlica-Wagner et al. 2018). The redMaPPer SDSS DR8 catalogue contains 26 311 clusters over the redshift range $0.08 < z < 0.55$ (Rykoff et al. 2014), while the redMaPPer DES SV contains 786 clusters from $0.2 < z < 0.9$ (Rykoff et al. 2016) and the redMaPPer DES Y1 contains over 76 000 clusters from $0.1 < z < 0.8$ (McClintock et al. 2019).

The Wavelet Z Photometric (WaZP) (Aguena et al. 2021) algorithm has also been applied to DES Y1 data. WaZP uses an overdensity-based approach to finding clusters and provides membership probabilities for cluster members within a redshift slice, with some similarities to the algorithm described in this paper. However, WaZP does not incorporate stellar mass into the cluster detection process; instead it identifies overdensities exclusively from photometric redshifts and galaxy positions on the sky.

Another well-tested cluster-finding algorithm was developed by Wen, Han & Liu (2009), henceforth WHL. This algorithm has been applied to SDSS (Wen, Han & Liu 2012; Wen & Han 2015), Subaru Hyper Suprime-Cam [data: Aihara et al. (2018) and clusters: Wen & Han (2021)], and the Dark Energy Survey [data: The Dark Energy Survey Collaboration (2005) and clusters: Wen & Han (2022)]. The WHL algorithm, just like the algorithm described in this paper, uses the 'nearest-neighbour' technique to find neighbouring galaxies, but it uses a deterministic slice in redshift rather than the probabilistic approach developed in this work.

A different approach to galaxy cluster-finding is exemplified by the Adaptive Matched Identifier of Clustered Objects (AMICO) algorithm (Bellagamba et al. 2018), which uses the matched-filter technique to achieve probabilistic cluster detection and membership probabilities for individual galaxies in the cluster. This algorithm is able to detect galaxy clusters in the low-signal-to-noise regime, but this ability comes at the expense of computational efficiency. The AMICO algorithm has been applied to the third data release (de Jong et al. 2017) of the Kilo-Degree Survey (de Jong et al. 2013), resulting in a catalogue of 7988 candidate galaxy clusters (Maturi et al. 2019). Since we do not use the Kilo-Degree Survey data in this work, we do not perform a direct comparison with this algorithm.

The current largest galaxy cluster catalogue (Zou et al. 2021) was created by applying the Clustering by Fast Search and Find of Density Peaks (CFSFDP) algorithm (Rodriguez & Laio 2014) to the Dark Energy Spectroscopic Instrument (DESI) Legacy Survey Data Release 8 (Dey et al. 2019). This work identified 540 432 galaxy cluster candidates reaching $z=1$ in the Dark Energy Spectroscopic Instrument Legacy Survey. Zou et al. (2021) use Sunyaev-Zel'dovich and X-ray data to calibrate total galaxy masses from richness–mass relations for this sample. Similarly to WHL, Zou et al. (2021) use a deterministic redshift slice, rather than a probabilistic approach. As our work uses a similar data set from the DESI Legacy Surveys, we use this catalogue as the main benchmark for comparison throughout this paper.

In this paper we present a new galaxy cluster-finding algorithm, which we call CluMPR (Clusters from Masses and Photometric Redshifts). This algorithm is designed to minimize the impact of projection effects caused by uncertainties in the photometric redshifts (photo-$z$'s) of individual galaxies. In this work, we optimize the details of this algorithm to enable the selection of a high-purity sample of high-stellar-mass clusters. These choices aid in selecting those galaxy cluster candidates which are most likely to cause strong lensing and help us to optimize searches for transients associated with quiescent galaxies, while simultaneously minimizing contamination of the sample by lower mass galaxy groups. In particular, our design choices tend to favour clusters with a high central density, which optimizes the lensing efficiency of the clusters. Our algorithm determines total cluster stellar masses probabilistically by incorporating photometric redshift uncertainties and compensating for redshift-dependent mass incompleteness. We also provide a catalogue of cluster member galaxies and a catalogue of candidate gravitationally lensed quasars. Our catalogues are intended to complement existing galaxy cluster catalogues such as the ones created by Zou et al. (2021) and Yang et al. (2021) using DESI Legacy Survey data, and can also be used to compare the results of applying different algorithms to similar data.

This paper is organized as follows: In Section 2, we describe the data products we used to generate our galaxy cluster catalogue. Section 3.1 describes the CluMPR algorithm in detail. In Section 3.2, we describe our process for calibrating galaxy cluster parameters and assigning galaxy cluster member galaxies. In Section 4, we provide some summary graphs and data summarizing the characteristics of the CluMPR galaxy cluster catalogue. Finally, in Section 5 we summarize the results, applications, and potential future extensions of our work. The Data Availability section provides the publicly accessible locations of all data and code related to this paper, including the cluster catalogues and candidate lensed quasar catalogues. Appendices A and B describe details regarding the calibration of our algorithm; Appendix C describes a set of candidate gravitationally lensed quasars we have identified; and Appendix D describes the columns of all catalogues assembled in this work. Throughout this paper, we assume a cosmology with $H_0 = 69.6 \, \mathrm{km \, s^{-1} \, Mpc^{-1}}$, $\Omega_m = 0.286$, and $\Omega_\lambda = 0.714$ (Bennett et al. 2014).

## 2 DATA

The DESI Legacy Imaging Surveys (Dey et al. 2019) has produced multiband images of over a third of the sky (roughly 19 700 deg$^2$) in three optical bands ($g$, $r$, $z$) and two infrared bands from *WISE* (*W*1, *W*2) (Wright et al. 2010). The optical component of this survey is composed of three separate programmes: the Beijing-Arizona Sky Survey (BASS) (Zou et al. 2017), the Dark Energy Camera Legacy Survey (DECaLS) (Flaugher et al. 2015; Dey et al. 2019) combined with additional data from the Dark Energy Survey (The Dark Energy Survey Collaboration 2005), and the Mayall $z$-band Legacy Survey (MzLS) (Silva et al. 2016). A full description of the Legacy Imaging Surveys is available at Dey et al. (2019). The primary purpose of the DESI Legacy Imaging Surveys is to enable the selection of targets for spectroscopic follow-up with DESI (Dark Energy Spectroscopic Instrument; DESI Collaboration 2022). In this work, we used the DESI Legacy Survey Data Release 9 (Schlegel et al. 2021). Throughout this text, a 'sweep' refers to a continuous subsection of DESI Legacy survey photometry data (covering a rectangular area in RA/DEC) which is contained in a single file in the released data products.

Our analysis relies on the photometric redshift measurements described in Zhou et al. (2023). That work used the AB magnitude system and applied a correction for galactic extinction based on the extinction maps by Schlegel, Finkbeiner & Davis (1998) with correction from Schlafly & Finkbeiner (2011). We also utilize the stellar mass estimates for Legacy Survey objects from Zhou et al. (2023). The photometric redshifts and stellar masses were both estimated by applying a Random Forest algorithm trained to predict a given quantity from Legacy Survey photometry. The photometric redshifts were trained using a carefully vetted set of spectroscopic redshifts from previous surveys, while the stellar masses are trained to match the stellar population synthesis-based mass estimates from Stripe82-MGC (Bundy et al. 2015). Uncertainty estimates for the photometric redshifts were determined from the spread in the predicted redshifts for an object when repeatedly perturbing the original photometry (sampling from its estimated errors). There is evidence that these photo-$z$ errors may be systematically overestimated for luminous red galaxies (Zhou et al. 2021), but we do not attempt to correct for that effect in this work. In the remainder of this paper, we assume that all uncertainties in photo-$z$'s are Gaussian distributed; while this does hold true for the Luminous Red Galaxy (LRG) sample (Zhou et al. 2021), this has not been investigated for bluer galaxies. However, we expect blue galaxies to be relatively rare in clusters below $z = 1$ (Schechter 1976; Dahle et al. 2013), so deviations from this should have small impact on our work. More detailed descriptions of the methods and catalogues for determining the photo-$z$'s and stellar masses used here can be found in Zhou et al. (2021, 2023).

We apply several photometric cuts to select galaxies for our analysis which should have reliable estimates of both photometric redshifts and stellar masses and with as little contamination from stars as possible. We restrict ourselves to objects with:

(i) $z$-band magnitude $<21$, as the Legacy Survey photo-$z$'s are less well-constrained at $z > 21$, as described in Zhou et al. (2021),

(ii) DESI legacy Survey DR9 bitmask value of 0 or 4096, corresponding respectively to unmasked galaxies and Siena Galaxy Atlas objects (Moustakas et al. 2023).[1]

(iii) Either *Gaia* photometric mean magnitude $> 19$, *Gaia* astrometric excess noise $>10^{0.5}$, or *Gaia* astrometric excess noise $= 0$; this corresponds to the *Gaia* PSF Criterion for excluding stars as used in the DESI Bright Galaxy Survey target selection, cf. Zarrouk et al. (2022),

(iv) TYPE (which indicates the type of model which provided the best fit in the Tractor DR9 photometry) not equal to 'PSF'; this effectively removes stars and quasars while keeping the vast majority of galaxies, since bright galaxies are detectably extended, and

(v) Photometric redshift $>0.01$; this eliminates low-$z$ objects, for which both the photometric redshifts and the deblending of galaxy images are less reliable.

We also restrict ourselves to only those galaxies which have a stellar mass estimated within Zhou et al. (2023). This will exclude those objects which do not have valid photometry in at least one of the $g$, $r$, or $z$ bands (W1 and W2 magnitudes have no effect) as well as those that do not satisfy a stellar rejection cut of $r - W1 > 1.75 \times (r - z) - 1.1$.

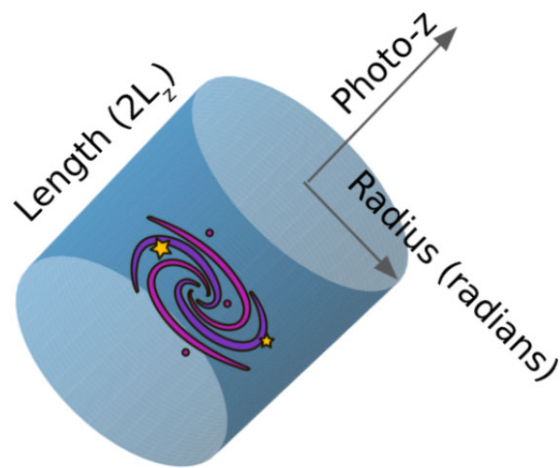

**Figure 1.** Search cylinder for CluMPR algorithm, oriented lengthwise along redshift (line-of-sight) direction and centred on the potential cluster centre galaxy.

Due to these cuts as well as our requirement of $z$ magnitude $<21$, the resulting galaxy sample is highly uniform across the sky, with minimal impact from depth variations in the Legacy Survey imaging [which is complete to $z > 22$ everywhere within the footprint (Dey et al. 2019)].[2] To account for the edges of sweeps, we include galaxies from all neighbouring regions that are within 0.285 degrees of the edges of a given sweep when running the algorithm (but we only keep those clusters that are within the boundaries of the sweep of interest to prevent duplicates). To handle the edges of the DR9 footprint, we flag clusters located within 0.285 degrees of the edge in our catalogues. Similarly, to handle effective holes in the footprint introduced by foreground galaxies, we flag clusters near large galaxies (see Section 3.1.5). We do not account for the edges of holes in the footprint due to bright stars.

## 3 METHODS

### 3.1 CluMPR algorithm overview

The CluMPR algorithm is comprised of five essential steps:

(i) Potential cluster centre selection: Select galaxies with large stellar mass.

(ii) Membership probability calculation: For each potential cluster centre galaxy, evaluate the expected number of neighbouring galaxies within a cylinder oriented lengthwise along the line of sight ($z$) direction (see Fig. 1). In the direction of the cylinder's radius, membership is determined deterministically using a binary search tree. Along the $z$ direction, membership is determined probabilistically based on photo-$z$ uncertainties.

(iii) Richness thresholds: Keep potential cluster centre galaxies with a significant excess of neighbouring galaxies (this is equivalent to defining a galaxy cluster by setting a minimum richness).

(iv) Aggregation: Neighbouring potential cluster centre galaxies are aggregated and the galaxy with the highest local stellar mass is chosen as the cluster centre.

(v) Cluster parameter correction: Biases in cluster parameters are estimated from an initial run of the algorithm. The algorithm is then

---

[1] For more details on DR9 bitmasks, see here.

[2] See also the DESI Legacy Survey DR9 depths provided here.

run again with a probabilistic reweighting and correction scheme that accounts for these biases.

This cylinder-based method is a probabilistic variant of the widely used (deterministic) 'counts-in-cylinders' technique (Hogg et al. 2004; Kauffmann et al. 2004; Blanton et al. 2006; Barton et al. 2007; Berrier et al. 2011). The individual steps of the CluMPR algorithm are explained in greater detail below.

### 3.1.1 Potential cluster centre selection

Potential cluster centre galaxies are chosen from the general pool of galaxies based on a minimum stellar mass threshold. We set this minimum stellar mass threshold to be log(stellar mass) > 11.2 [log($M_\odot$)]. This mass threshold was chosen based on visual inspection of galaxy clusters (we chose to minimize the fraction of low-mass groups and spurious detections in order to create a high-purity cluster catalogue, but this mass threshold can be lowered if a lower purity catalogue is desired).

### 3.1.2 Membership probability calculation

For each potential cluster centre galaxy chosen in 3.1.1, we count the number of neighbours in a cylinder oriented lengthwise along the $z$ direction and centred on the potential cluster centre galaxy.

First, galaxies are selected within a fixed physical radius of the cluster centre galaxy (we use three radii: 1, 0.5, and 0.1 Mpc. The significance of these radii is explained in the subsequent sections). This step is accomplished using a binary search tree; our implementation uses the cKDTree algorithm from the SCIPY package (Virtanen et al. 2020). In order to avoid propagating photo-$z$ errors into the angular positions of galaxies, we perform this step using angular coordinates. The physical radius is calculated in angular coordinates using the following equation:

$$R_{\rm ang}(z) = (R_{\rm phys}) \frac{1+z}{D(z)}, \quad (1)$$

where $R_{\rm ang}$ is the radius in angular coordinates (radians), $R_{\rm phys}$ is the physical radius in megaparsecs, $z$ is the redshift, and $D(z)$ is the comoving distance given by

$$D(z) = \int_0^z \frac{dz'}{H(z')}, \quad (2)$$

where $H(z')$ is the Hubble–Lemaitre parameter. A graph of equation (1) for $R_{\rm phys} = 1\,{\rm Mpc}$ is shown in Fig. 2. For each potential cluster centre galaxy, we calculate the radius using equation (1) and the photo-$z$ of the cluster centre galaxy. To ensure equal area projection of angular coordinates, we use a sinusoidal (Sanson–Flamsteed) projection. To avoid issues associated with very small values for angular coordinates, we add an offset of 50 radians to each coordinate.

Next, to determine cluster membership along the z-direction (lengthwise along the cylinder), we apply a redshift cut based on the photo-$z$ of the cluster centre galaxy. The length of this cut (the half-length of the cylinder) is given by

$$L_z(z) = 2\overline{\sigma_z}, \quad (3)$$

where $\overline{\sigma_z}$ is the interpolated redshift-dependent binned average of the photo-$z$ error for all galaxies. To train our model for $\overline{\sigma_z}$, we used 100 bins between the lowest and highest photo-$z$'s in the training sample; we limited $\overline{\sigma_z}$ to 0.1 at high z. We used 20 randomly selected DESI Legacy Survey sweep files [10 from North (BASS/MzLS) and

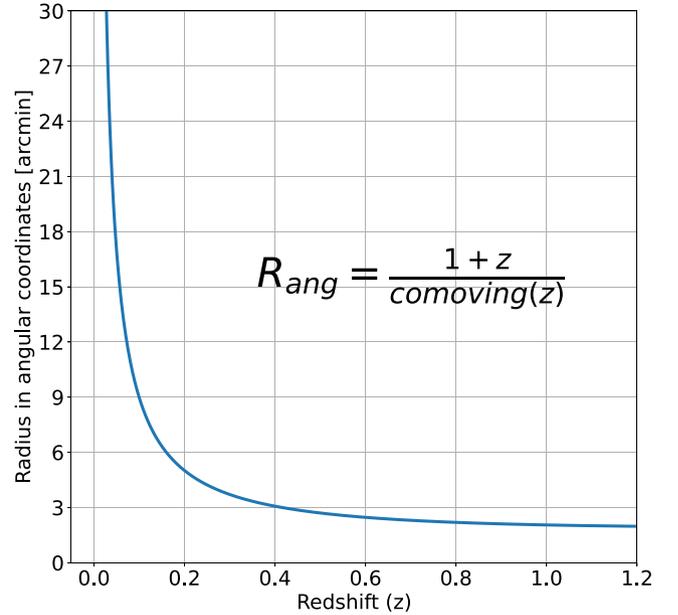

**Figure 2.** Search cylinder radius in angular coordinates (radians) as a function of redshift. Obtained from equation (1).

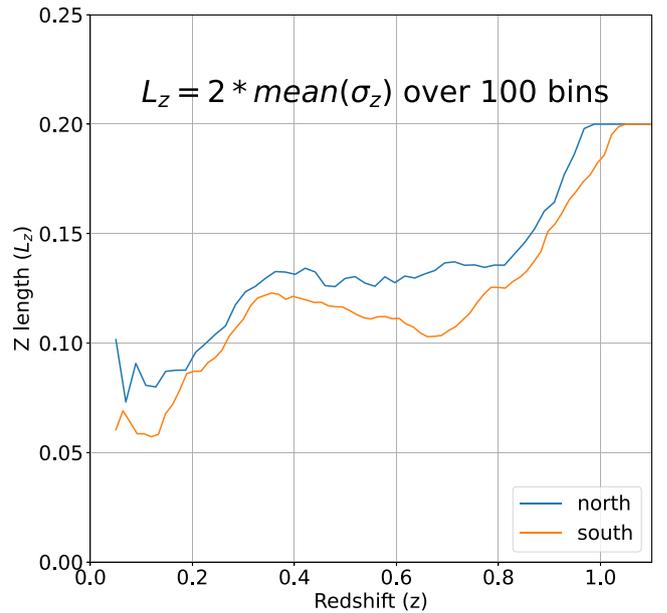

**Figure 3.** Interpolated redshift-dependent binned average of the photo-$z$ error as a function of redshift. This is the half-length of the search cylinder. Note that $L_z$ is limited to 0.2. South refers to the DECaLS survey data, while north refers to the data from BASS and MzLS.

10 from South (DECaLS)] as our training samples. The linearly interpolated trained results for $\overline{\sigma_{z,{\rm north}}}$ and $\overline{\sigma_{z,{\rm south}}}$ are shown in Fig. 3. $L_z$ is significantly larger than the true redshift extent of a galaxy cluster at all redshifts, which means that a probabilistic approach to cluster membership along the z direction is required.

We use $L_z$ to determine the probability that a galaxy within the search cylinder's radius is a true member of the cluster (that is, the probability that it is truly within the cylinder). In order to determine this probability, we use statistical resampling, whereby we create many realizations of each galaxy's potential true redshift. Since the photometric redshift catalogue we are using provides only Gaussian uncertainties on for each object's redshift, these realizations are drawn from a normal distribution centred on the photo-$z$ of the galaxy and with a standard deviation equal to the photo-$z$ error for the galaxy. The membership probability for an object is then the fraction of its redshift realizations that fall within $L_z$ of the cluster centre galaxy's photo-$z$ (before applying any reweighting; see Section 3.2.1 for details of the reweighting process).

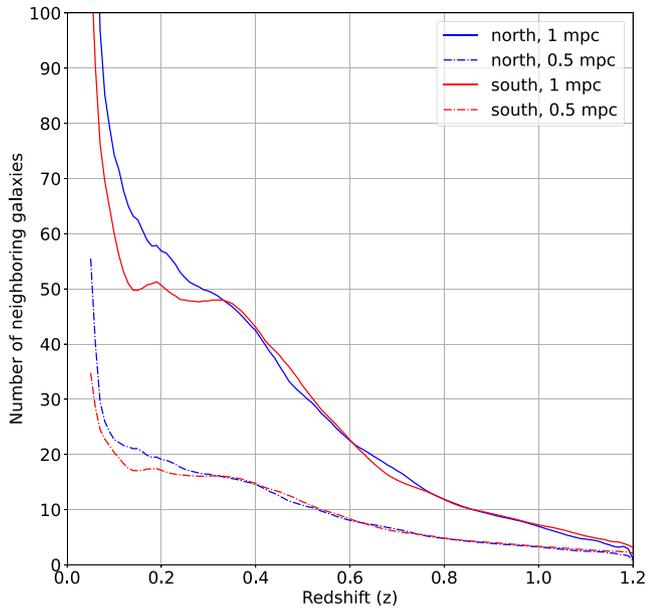

**Figure 4.** Interpolated redshift-dependent thresholds for 1-Mpc and 0.5-Mpc cylinders. South refers to the DECaLS survey data, while north refers to the data from BASS and MzLS.

### 3.1.3 Richness thresholds

In order to identify galaxy clusters among the potential cluster centre galaxies selected in Step 3.1.1, we apply a threshold for the minimum number of galaxies that must be located within the galaxy's search radius. This is equivalent to defining what is a galaxy cluster by setting a minimum richness threshold. We apply a redshift-dependent threshold to both the 1-Mpc cylinder and the 0.5-Mpc cylinder; only potential cluster centre galaxies that satisfy both thresholds are kept as galaxy clusters. For the 1-Mpc cylinder, the threshold we used is

$$T_1 > \overline{N} + 1.8\sigma_N, \qquad (4)$$

and for the 0.5-Mpc cylinder, the threshold we used is

$$T_{0.5} > \overline{N} + 1.2\sigma_N, \qquad (5)$$

where $\overline{N}$ is the redshift-binned average number of galaxies within a cylinder of that size, and $\sigma_N$ is the redshift-binned standard deviation of the number of galaxies within a cylinder of that size. These thresholds were chosen based on visual inspection for purity of resulting galaxy clusters. We assume a Poisson distribution for the number of galaxies within a cylinder, which allows us to define

$$\sigma_N = \sqrt{\overline{N}}. \qquad (6)$$

The richness threshold is interpolated linearly as a function of $z$; the results of this interpolation are shown in Fig. 4. The training data for these thresholds are the same as in Section 3.1.2, and the redshift bins are $z=0.01$ apart with a width of $z = 0.05$, ranging from the lowest to the highest photo-$z$ in the training samples.

### 3.1.4 Aggregation

In order to remove duplicate galaxy clusters (several centre galaxy candidates for one galaxy cluster), we use a similar method to Section 3.1.2, except that the radius of the aggregation cylinder is 1.5 Mpc, and the $L_z$ cutoff is not done probabilistically.

First, all cluster central galaxies are sorted by (uncorrected, see 3.2.2) total stellar mass within a 0.5 Mpc radius. Beginning with the cluster central galaxies with the greatest 0.5 Mpc stellar mass, neighbouring cluster central galaxies within $R = 1.5$ Mpc and $L_z$ are labelled as belonging to the same cluster. The galaxy with the greatest (uncorrected) total stellar mass within a 0.1 Mpc radius is then chosen as the cluster centre galaxy.

### 3.1.5 Data cleaning

We apply several flags to our galaxy clusters to remove poor objects. First, galaxy clusters within 0.285 degrees of the edge of the DESI Legacy Survey footprint are flagged. This flagging is performed in three steps. First, we create a histogram map of detected galaxy cluster positions and set all non-zero values to one. Next, we perform a close operation (dilate + erode with a circular kernel of 10 pixels, which in this context is roughly 0.32 by 0.64 degrees for the minor and major axes of the kernel in coordinate space) in order to fill in space between the clusters; this effectively creates a histogram map of the DESI Legacy Survey footprint. Finally, we erode the map by an oval kernel which (in coordinate space) corresponds to a circle with a 0.285 degree radius; we then flag all clusters whose positions are outside this eroded map. We also flag clusters within 5 arcmin of large foreground galaxies from the Siena Galaxy Atlas (SGA) (Moustakas et al. 2023). Our criteria for the SGA galaxies are either

(i) D26 > 1.5,
(ii) R_MAG_SB24 < 13.5, R_MAG_SB24 > 0, and Z_LEDA < 0.05

Some of these flagged objects are false detections (which are mostly due to failed de-blending of foreground galaxies), while others are real galaxy clusters with significant foreground contamination which could affect the photo-$z$ and stellar mass estimates. Finally, we remove all clusters located in the isolated strips of the DESI Legacy Survey footprint with declination $< -10.3$ degrees and $110 <$ Right ascension (degrees) $< 250$. We provide a catalogue which includes all flagged objects as a supplement to the main CluMPR catalogue.

### 3.2 Cluster parameters

Each galaxy cluster in our catalogue is labelled with a unique 'gid', which is created from the cluster centre galaxy's coordinates using the following formula:

round(RA, 6 decimals) $* 10^{16}$
   +round(DEC + 90, 6 decimals) $* 10^{6}$,

where RA is right ascension and DEC is declination (both in degrees). Each galaxy cluster has a richness estimate (number of neighbours) for the 1, 0.5, and 0.1 Mpc search radii. The 'gid' is necessary to identify the galaxy cluster members described in Section 3.2.3. Additionally, we provide galaxy cluster parameters for redshift and mass as described below.

*3.2.1 Probability and mass weighting of cluster redshift parameters*

For each galaxy cluster we calculate the following redshift parameters (in addition to providing the photo-$z$ information for the cluster centre galaxy):

(i) the average photo-$z$ of all galaxy cluster member galaxies;

(ii) the average photo-$z$ of all galaxy cluster member galaxies (weighted by reweighted membership probability);

(iii) the average photo-$z$ of all galaxy cluster member galaxies (weighted by galaxy stellar mass times reweighted membership probability);

(iv) the standard error in the redshift of the cluster, computed from the photo-$z$'s of all galaxy cluster member galaxies;

(v) the standard error in the redshift of the cluster, computed from the photo-$z$'s of all cluster member galaxies weighted by their reweighted membership probabilities; and

(vi) the standard error in the redshift of the cluster, computed from the photo-$z$'s of all cluster member galaxies weighted by their stellar masses times their reweighted membership probability.

For the standard error, we use the unbiased estimate for the weighted standard error:

$$S_{\bar{x}} = \frac{S_x}{\sqrt{n_{\text{eff}}}}, \quad (7)$$

where $S_x$ is the (reweighted) membership probability-weighted standard deviation. $n_{\text{eff}}$ takes into account the effect of the weighting on degrees of freedom and is defined as

$$n_{\text{eff}} = \frac{(\sum_i w_i)^2}{\sum_i (w_i^2)}, \quad (8)$$

where $w_i$ is the reweighted membership probability for the $i$-th galaxy.

We found that if potential member galaxies of all redshifts are given a weight that depends only on the magnitude of the deviation from the estimated cluster's redshift but not its sign, we obtain photo-$z$ estimates for the clusters which are biased high for galaxy clusters at low z and biased low for galaxy clusters at high z. This appears to be a consequence of an Eddington bias (Eddington 1913) caused by the higher abundance of non-member galaxies that fall within the cylinder of a given cluster by chance at redshifts where the number of galaxies in our sample is higher; since galaxies between $0.4 < z < 0.6$ are most numerous in our sample, this tends to positively bias the redshifts of clusters at lower $z$ and negatively bias the redshifts of clusters at higher $z$ than this. This effect is further compounded by the larger average photo-$z$ errors for higher redshift galaxies.

We correct for this non-linear effect by running the cluster-finding algorithm twice; the results of a first run are used to calculate a reweighting that is then applied to objects during the second run. During the second run, we still use the non-reweighted richness and stellar mass to identify clusters in the same way as during the initial run (to ensure that the identified galaxy clusters do not change significantly between runs and to eliminate the need for recalculating the richness thresholds). The reweighting factor applied to each galaxy is calculated as follows:

(i) The galaxy cluster membership catalogue (which contains the probability of each galaxy being a cluster member) is binned in redshift. In this work, we use 20 bins covering the range from $z = 0.1$ to $1.0$.

(ii) Next, within each bin, we calculate a normalized histogram of '$\Delta z$' values, where $\Delta z$ is the difference between a potential member galaxy's redshift and the best estimate of the cluster redshift. We found the cluster central galaxy's median photo-$z$ to be a sufficiently accurate proxy for the cluster spectroscopic redshift for this purpose. Hence, we define $\Delta z$ as the difference between a galaxy's redshift and the median photo-$z$ of the central galaxy of the cluster it is a potential member of.

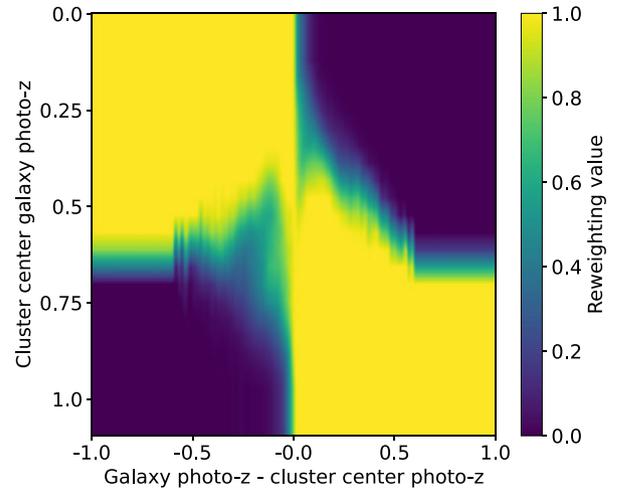

**Figure 5.** Visualization of values of the reweighting factor applied to correct Eddington bias, shown as a function of the difference between the galaxy photo-$z$ and cluster photo-$z$ (corresponding to the x axis) and the cluster photo-$z$ itself (corresponding to the y axis). One could interpret the plotted quantity as the relative values of an unnormalized prior for $\Delta z$, shown at different possible values of $z$.

(iii) The resulting histograms have a peak at $\Delta z > 0$ for low-$z$ bins and a peak below at $\Delta z < 0$ for high-$z$ bins. Next, we flip the histogram symmetrically around $\Delta z = 0$ and then divide the original histogram's values by the values of the flipped histogram. Finally, we set all values greater than one to one. To reduce variation caused by binning, we apply a $3 \times 3$ pixel box kernel convolution and then perform a linear interpolation to obtain the reweighting at any arbitrary at $\Delta z$ value (with nearest neighbour extrapolation beyond $|\Delta z| = 0.6$).

The result of this procedure (illustrated in Fig. 5) is a value between 0 and 1 which we multiply by each galaxy's membership probability (determined as described in Section 3.1.2) to obtain a reweighted probability that an object is a cluster member. By flipping the histogram around $\Delta z = 0$ (and limiting values to be less than 1), we force the reweighted redshift distribution of cluster members about the cluster centre galaxy to be symmetric. This procedure has the added benefit of mitigating any potential effects from the increase of the amount of volume over which galaxies may be interlopers as redshift increases. One could interpret this reweighting scheme in a Bayesian sense as an unnormalized (improper) prior which we learn from the observed Eddington bias in the first run of our algorithm and then combine with our prior knowledge that member galaxies should be distributed roughly symmetrically in redshift space about the cluster centre galaxy.

*3.2.2 Cluster richness and stellar mass estimates*

We provide a background-corrected richness estimate which is background-corrected but not corrected for survey incompleteness. This richness is calculated by summing the reweighted membership probabilities of the galaxy cluster member galaxies and then subtract-ing the average galaxy background from this sum. The calculation of the average galaxy background is described in Appendix A.

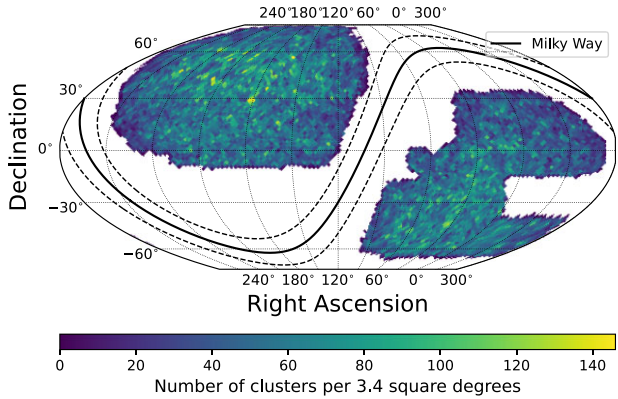

**Figure 6.** Density of galaxy clusters found by the CluMPR algorithm in the DESI Legacy Survey on the sky.

We determine galaxy cluster stellar masses by summing the reweighted membership probability-weighted stellar masses of all cluster member galaxies within the 1, 0.5, and 0.1 Mpc search radii. We exclude galaxies with a mass below a threshold described in Appendix B (equation B3). We then subtract the average galaxy stellar mass background, which is described in Appendix A. Finally, we apply a redshift-dependent correction factor which compensates for the bias due to redshift-dependent mass incompleteness, commonly known as Malmquist bias (Malmquist 1922, 1925). This factor is described in Appendix B; the result is an incompleteness-corrected cluster stellar mass containing all cluster members above $\log(M_\star/M_\odot) = 10$.

*3.2.3 Cluster membership*

For every galaxy cluster, we have compiled a list of member galaxies (galaxies whose membership probability > 0.0027). For galaxies which are found to belong to more than one galaxy cluster, we assign the galaxy cluster for which there is the highest probability of membership. For each galaxy cluster, the cluster member galaxies can be identified using the 'gid' parameter (see Section 3.2). The membership probability for the cluster member galaxies provided in the membership catalogue is not the same as the probability defined in Section 3.1.2. Instead, it is the reweighted probability of the member galaxy having a true redshift equal to the photo-$z$ of the galaxy cluster's central galaxy, assuming a Gaussian distribution centred on the member galaxy's redshift with a standard deviation equal to that member galaxy's photo-$z$ error.

## 4 RESULTS

The CluMPR algorithm has returned 298 487 candidate galaxy clusters from $0.1 < z < 1$ (excluding flagged objects). Galaxy clusters with $z < 0.1$ are excluded as our algorithm is not optimized for those objects, which are frequently contaminated by poorly deblended foreground galaxies. The distribution of our catalogue's galaxy clusters on the sky is shown in Fig. 6. Fig. 7 shows a few examples of galaxy clusters from our catalogue at various redshifts. Fig. 8 shows the redshift and stellar mass distribution of CluMPR clusters. For $0.1 < z < 0.3$, the sample is volume-limited; above $z = 0.3$, the sample is limited by incompleteness. The stellar mass correction factor causes the stellar mass distribution to remain relatively flat across $z$. Low-mass galaxy clusters are excluded by our thresholds and incompleteness at high $z$, while high mass galaxy clusters are volume-limited. The rise in high-stellar-mass clusters at $z < 0.3$ indicates that the catalogue is volume-limited for those redshifts; the decline in stellar mass at $z > 0.8$ is likely to be driven by systematic photo-$z$ biases which lead to underestimated distances to individual galaxies at $z > 0.8$ (see Fig. 13 for details). A first order correction to this bias can be calculated by fitting a model for the bias in photometric redshifts relative to available spectroscopic redshifts. Systematics in the galaxy stellar masses or cosmological evolution may also play a role in the decline in total stellar mass at high redshift.

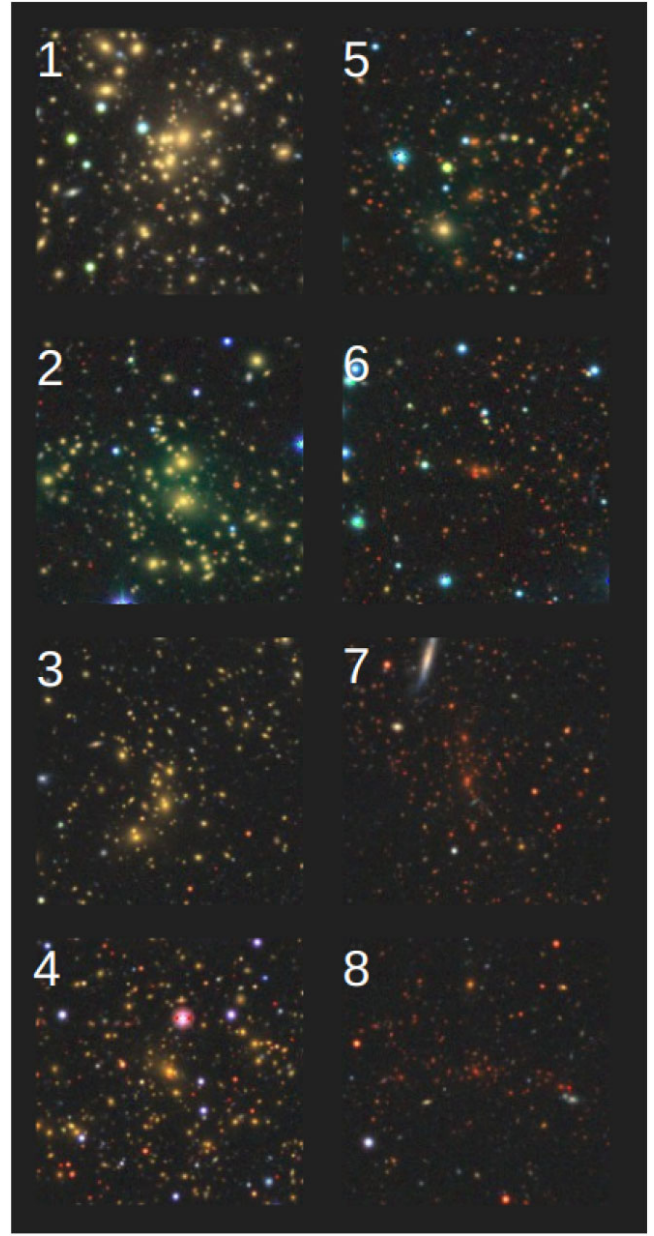

**Figure 7.** Examples of galaxy clusters found by the CluMPR algorithm in the DESI Legacy Survey imaging. The central galaxy redshifts and CluMPR gid names of the clusters are as follows: 1. J1978752910088654862, $z = 0.193$ 2. J1394763660141732540, $z = 0.258$ 3. J35853440059605682, $z = 0.320$ 4. J2629163750112866265, $z = 0.404$ 5. J1093980870127751457, $z = 0.516$ 6. J2692212310130135307, $z = 0.602$ 7. J3518636620087924510, $z = 0.733$ 8. J1642499610086373541, $z = 0.826$.

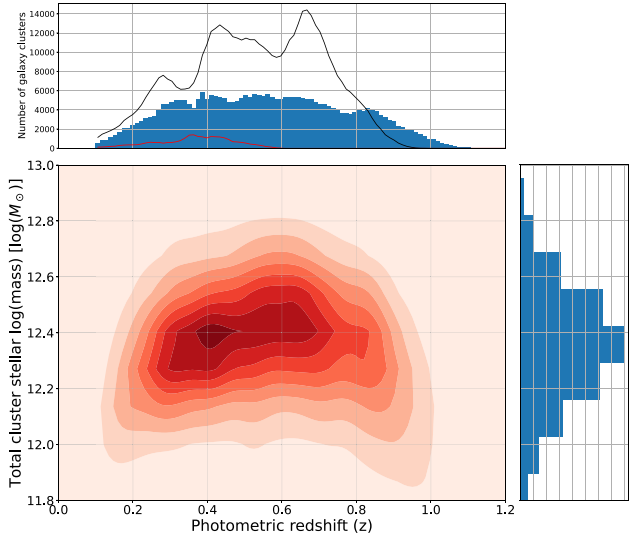

**Figure 8.** 1-Mpc total stellar mass vs photo-$z$ for galaxy clusters found by the CluMPR algorithm in the DESI Legacy Survey imaging. Cluster stellar masses have been background-subtracted and corrected for incompleteness with stellar mass correction factor. The top histogram shows the redshift distribution of CluMPR clusters with the redshift distributions of Zou et al. (2021) (black, upper line) and redMaPPer (red, lower line) overplotted. The right side histogram shows the stellar mass distribution of CluMPR clusters.

### 4.1 Cross-matching with Zou et al. (2021)

The galaxy cluster catalogue described in Zou et al. (2021) was compiled using the same DESI Legacy Survey data as our catalogue, so cross-matching the catalogues yields a direct comparison of our implementation of CluMPR with the implementation of the CFSFDP algorithm by Zou et al. (2021). For cross-matching, we apply a simple 2D cross-matching which ignores line-of-sight coincidence. The radius for determining a match was 5 arcmin, which is sufficient to handle most differences in the choice of cluster centre galaxy while minimizing spurious matches. We found that this radius worked reasonably well at all redshifts.

Overall, Zou et al. (2021) found 84.6 per cent of the galaxy clusters identified by CluMPR above $z = 0.1$. This high match rate indicates that the CLuMPR sample contains relatively high-mass and high-richness galaxy clusters which are not subject to significant threshold effects when used as the target for matching Zou et al. (2021) clusters (Bahcall et al. 2003). Fig. 9 shows the stellar mass within 1 Mpc for CluMPR galaxy clusters as a function of redshift; colour indicates the fraction of CluMPR clusters found by Zou et al. (2021). Between $z = 0.1$ and $z = 0.75$, there is broad agreement between CluMPR and Zou et al. (2021) on high-richness objects, but Zou et al. (2021) miss some low-richness candidates identified by CluMPR. At $z > 0.75$, Zou et al. (2021) miss a large fraction of clusters identified by CluMPR, including high-mass clusters. Visual inspection of the high-$z$ clusters missed by Zou et al. (2021) indicates a high fraction of true galaxy clusters. The likely reason for Zou et al. (2021) missing these objects is that they use a minimum richness of 10 which is not corrected for incompleteness; at high redshift, many clusters will have a low observed richness. This effect is best illustrated by Fig. 10, which shows the 1-Mpc CluMPR richness as a function of redshift; colour indicates the fraction of CluMPR galaxies found by Zou et al. (2021). At low-$z$, the CluMPR richness threshold for a cluster is relatively high, but it drops rapidly with $z$; by $z$ 0.75, the richness threshold falls below 10, and the fraction of galaxy clusters recovered by Zou et al. (2021) falls to 0.5. CluMPR is able to maintain relatively high purity at high z despite the low richness threshold by using high-stellar mass galaxies as a reliable starting point for finding galaxy clusters, which works even in the absence of a large number of cluster member galaxies.

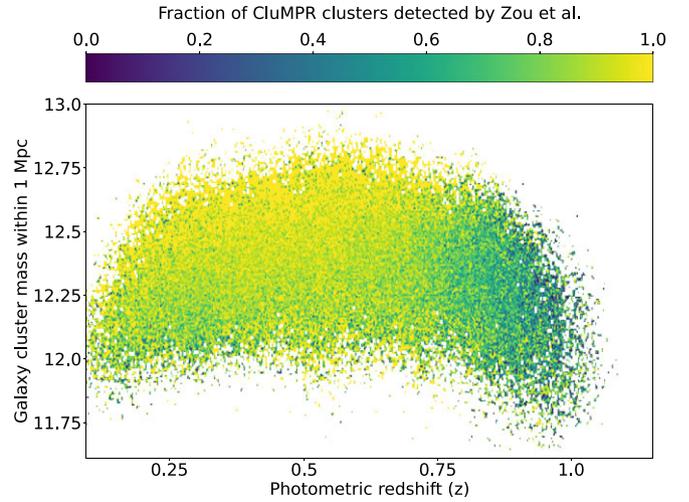

**Figure 9.** 1-Mpc total stellar mass versus photo-$z$ for galaxy clusters found by the CluMPR algorithm in the DESI Legacy Survey imaging (same as Fig. 8). Colour represents the fraction of CluMPR clusters found by Zou et al. (2021). The catalogue by Zou et al. (2021) has broad agreement on high mass clusters.

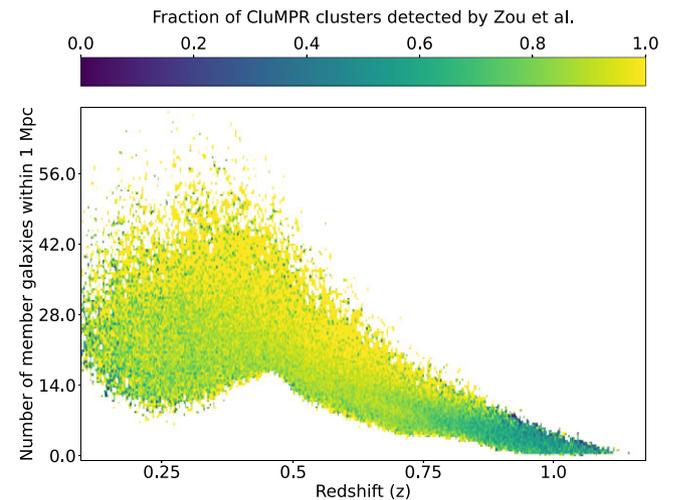

**Figure 10.** 1-Mpc richness (expected number of neighbours, as defined in Section 3.2.2) versus photo-$z$ for galaxy clusters found by the CluMPR algorithm in the DESI Legacy Survey imaging. Colour represents the fraction of CluMPR clusters found by Zou et al. (2021). The lower bound on richness for clusters with $z > 0.45$ is set by the thresholds in Section 3.1.3 (compare Fig. 4). For $z < 0.45$, the lower bound on richness is strongly affected by the reweighting of member probabilities, reducing the resemblance to Fig. 4.

Overall, CluMPR finds 58.0 per cent of the galaxy clusters identified by Zou et al. (2021); however, this percentage increases to 86.9 per cent if one limits the Zou et al. (2021) sample to clusters with richness > 30. This behaviour is expected (Bahcall et al. 2003), as the different types of thresholds (richness versus stellar mass) used by different algorithms and uncertainty in the quantities used as thresholds can combine to reduce the matching rate to values near 50 per cent if one catalogue has a lower threshold than the other.

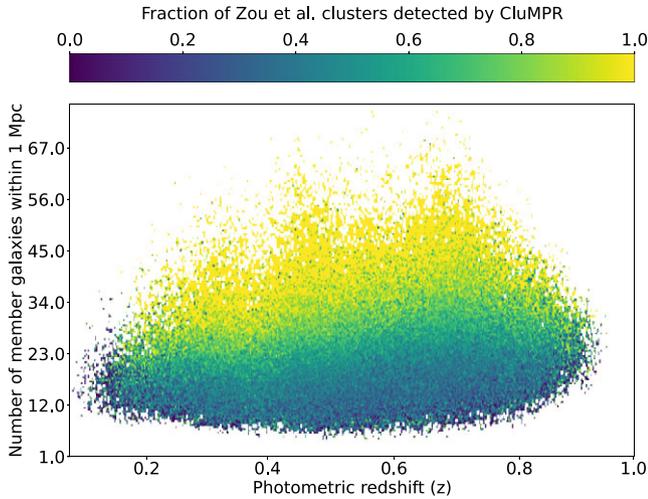

**Figure 11.** 1-Mpc richness (expected number of neighbours) versus photo-$z$ for galaxy clusters found by Zou et al. (2021) in the DESI Legacy Survey imaging. Colour represents the fraction of Zou et al. (2021) clusters found by CluMPR.

Hence, we focus on the percentage of Zou et al. (2021) clusters with richness > 30 which were identified in common, since by restricting to the most strongly detected objects we ensure that the threshold effects are negligible. Fig. 11 shows the 1 Mpc richness for Zou et al. (2021) galaxy clusters as a function of redshift; colour indicates the fraction of CluMPR clusters found by Zou et al. (2021). As in Fig. 9, there is broad agreement on high-richness objects. At all redshifts, CluMPR does not find most of the low-richness galaxy clusters identified by Zou et al. (2021). Thus, it is likely that the CluMPR sample is better optimized for detecting high-mass clusters with high purity (which aligns with our original goals). An implementation of CluMPR with lower richness thresholds would likely yield results more similar (for $z < 0.75$) to Zou et al. (2021).

### 4.2 Cross-matching with SDSS redMaPPer and SDSS spectroscopy

Both CluMPR and Zou recover most of the objects from the SDSS redMaPPer cluster catalogue. Using the 5 arcmin radius cross-matching described in Section 4.1, CluMPR recovers 97 per cent of redMaPPer clusters, while Zou et al. (2021) recover 99 per cent of redMaPPer clusters. Fig. 12 shows the 1-Mpc cluster stellar mass versus redMaPPer richness. As redMaPPer richness is known to be correlated with total cluster mass (Simet et al. 2017), the observed correlation between CluMPR stellar mass and redMaPPer richness is a strong indicator that CluMPR stellar masses are correlated with total cluster mass.

We provide spectroscopic redshifts for galaxies whose spectra have been obtained by the Sloan Digital Sky Survey (SDSS) (Blanton et al. 2017). Fig. 13 shows the residuals between SDSS spectra and three photometric estimates for cluster redshift from CluMPR, as well as the residuals for redshift estimates by redMaPPer and Zou et al. (2021). The bias for average redshifts obtained by CluMPR is caused by line-of-sight galaxies falling within the cluster; this bias is increased when photo-$z$ errors are systematically overestimated, as this creates an artificially long cylinder and increases the likelihood for galaxies to fall within the cylinder. Zhou et al. (2021) indicate that photo-$z$ errors are indeed overestimated, so this bias would be reduced with an improved redshift uncertainty quantification.

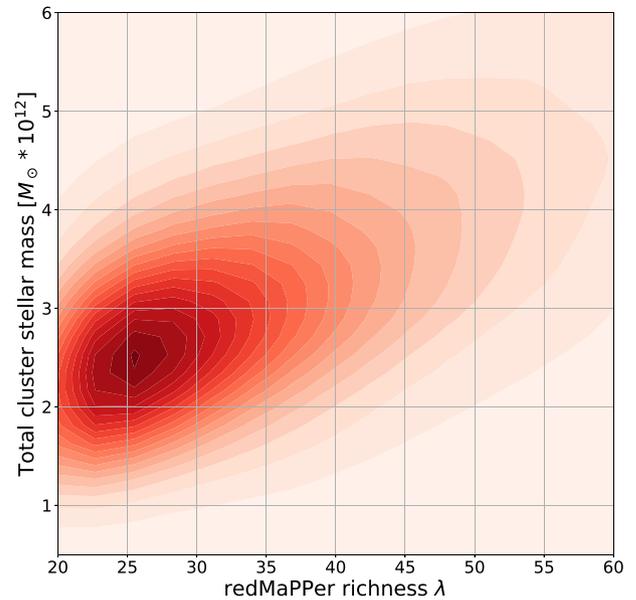

**Figure 12.** 1-Mpc cluster stellar mass versus redMaPPer richness. The redMaPPer catalogue does not have clusters below richness = 20. The cluster stellar mass has been background-subtracted and corrected for incompleteness with stellar mass correction factor. There is a clear correlation between CluMPR cluster stellar mass and redMaPPer richness, which indicates that CluMPR stellar masses are likely correlated with total cluster mass as well.

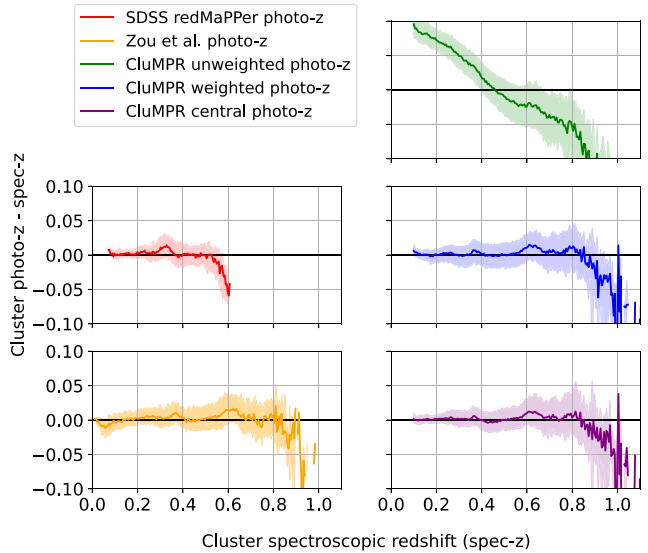

**Figure 13.** Residuals between SDSS spectroscopic redshifts and various estimates of cluster redshift from photo-$z$'s. Plotted intervals contain 68.2 per cent of clusters. Our catalogue offers several redshift estimates; the photo-$z$ of the cluster central galaxy is least biased (bottom right). The plotted weighted average (middle right) uses mass∗probability weighting, which reduces bias significantly compared to an unweighted average redshift (top right) and yields results with a bias similar in magnitude to that found when the cluster central galaxy's photo-$z$ is used to determine redshift. We include a catalogue of the member galaxies for all clusters, which can be used to calculate alternative estimates of the cluster redshift. All plotted redshift estimates underestimate spectroscopic redshift near the high-$z$ limit due to sharply decreasing galaxy counts at increasing redshifts (attenuation bias, see Freeman et al. 2009), which leads to a redshift estimate for the cluster central galaxy which is biased low.

However, our algorithm is able to significantly reduce the bias by weighting the cluster members by mass × reweighted probability (shown, middle right) or just reweighted probability (not shown, roughly equivalent in performance to mass × probability). This indicates that our stellar mass and richness estimates are likely also improved by our probabilistic approach.

## 5 CONCLUSIONS

We have used the CluMPR algorithm to identify nearly 300 000 galaxy clusters using the DESI Legacy Survey imaging data. This algorithm provides accurate photo-$z$ estimates, incompleteness-corrected stellar masses for each cluster, and cluster membership probabilities for potential member galaxies. Our catalogue shows good correspondence to previous catalogues such as Zou et al. (2021) and SDSS redMaPPer (Rykoff et al. 2014) in overlapping regimes, and visual inspection of the detected galaxy clusters indicates that our methods have produced a high-purity sample.

CluMPR has several advantages over non-probabilistic cluster-finding algorithms such as the WHL algorithm (Wen et al. 2009) or the CFSFDP algorithm (Rodriguez & Laio 2014) as used by Zou et al. (2021). As an example, for strong lensing searches, CluMPR enables us to select a sample of high-stellar-mass clusters that simultaneously has high purity. Higher redshift lensing searches are further enhanced as CluMPR is able to more easily identify galaxy clusters at higher $z$ ($z > 0.75$), a regime where smaller number of cluster galaxies will be brighter than magnitude thresholds and hence apparent richnesses will be lower. However, by using the presence of a single massive galaxy ($log(M_\star) > 11.2$) as a signpost to reliably anchor cluster assignments, CluMPR is able to still identify clusters with high purity even when the apparent richness drops. In contrast to red-sequence-based algorithms such as redMaPPer (Rykoff et al. 2014), CluMPR includes all potential galaxy cluster members in its search, which leads to a more complete member sample. This difference will become more relevant for deeper surveys, as we would expect a higher fraction of cluster members which are not on the red sequence at high $z$ (Schechter 1976; Dahle et al. 2013). Finally, our probabilistic approach to cluster membership reduces systematic errors in cluster properties caused by the incorporation of infalling foreground or background galaxies.

The CluMPR DESI Legacy Survey catalogue can now be used by astronomers to search for strongly lensed time-variable phenomena (such as supernovae and quasars) as well as other applications. As an example of how this catalogue may be used, we provide a supplementary catalogue of candidate lensed quasars in Appendix C. The nine best candidates identified from this sample include two known systems and a possible changing-look lensed QSO with SDSS spectroscopy. The CluMPR DESI Legacy Survey catalogue can also be used for galaxy evolution studies and to search for merging galaxy clusters.

The CluMPR sample could be made more useful for cosmological studies by calibrating the relationship between the cluster stellar masses we have measured and total cluster masses as derived from Sunyaev-Zel'dovich, X-ray, or weak lensing measurements. Future data sets from Rubin Observatory (LSST Science Collaboration 2009), Euclid Observatory (Euclid Collaboration 2022), the Nancy Grace Roman Telescope (Spergel et al. 2015), and other next-generation surveys will provide another application for the CluMPR algorithm. These projects should provide higher precision photometric redshifts and useful stellar mass measurements for objects several magnitudes fainter than the Legacy Survey objects studied here. As a result, these deep imaging surveys could extend CluMPR's reach to $z = 2$ and beyond.

## ACKNOWLEDGEMENTS

MJY-B would like to acknowledge his PhD advisors Yashar Hezaveh and Laurence Perreault-Levasseur for their patience and advice as he finalized the results presented in this paper. We also thank Eli Rykoff for valuable and insightful commentary during the peer review process. Additionally, we thank Gourav Khullar for a helpful discussion of cluster-lensed quasars.

The efforts of MJY-B were supported by three grants from the NASA Pennsylvania Space Grant Consortium in 2020 and 2021, as well as by the bourse d'excellence du centenaire from the Fonds du Centenaire of the Department of Physics at the Université de Montréal. The efforts of JAN were supported by grant DE-SC0007914 from the U.S. Department of Energy Office of Science, Office of High Energy Physics.

The Legacy Surveys consist of three individual and complementary projects: the Dark Energy Camera Legacy Survey (DECaLS; Proposal ID #2014B-0404; PIs: David Schlegel and Arjun Dey), the Beijing-Arizona Sky Survey (BASS; NOAO Prop. ID #2015A-0801; PIs: Zhou Xu and Xiaohui Fan), and the Mayall z-band Legacy Survey (MzLS; Prop. ID #2016A-0453; PI: Arjun Dey). DECaLS, BASS and MzLS together include data obtained, respectively, at the Blanco telescope, Cerro Tololo Inter-American Observatory, NSF's NOIRLab; the Bok telescope, Steward Observatory, University of Arizona; and the Mayall telescope, Kitt Peak National Observatory, NOIRLab. The Legacy Surveys project is honored to be permitted to conduct astronomical research on Iolkam Du'ag (Kitt Peak), a mountain with particular significance to the Tohono O'odham Nation.

NOIRLab is operated by the Association of Universities for Research in Astronomy (AURA) under a cooperative agreement with the National Science Foundation.

This project used data obtained with the Dark Energy Camera (DECam), which was constructed by the Dark Energy Survey (DES) collaboration. Funding for the DES Projects has been provided by the U.S. Department of Energy, the U.S. National Science Foundation, the Ministry of Science and Education of Spain, the Science and Technology Facilities Council of the United Kingdom, the Higher Education Funding Council for England, the National Center for Supercomputing Applications at the University of Illinois at Urbana-Champaign, the Kavli Institute of Cosmological Physics at the University of Chicago, Center for Cosmology and Astro-Particle Physics at the Ohio State University, the Mitchell Institute for Fundamental Physics and Astronomy at Texas A&M University, Financiadora de Estudos e Projetos, Fundacao Carlos Chagas Filho de Amparo, Financiadora de Estudos e Projetos, Fundacao Carlos Chagas Filho de Amparo a Pesquisa do Estado do Rio de Janeiro, Conselho Nacional de Desenvolvimento Cientifico e Tecnologico and the Ministerio da Ciencia, Tecnologia e Inovacao, the Deutsche Forschungsgemeinschaft and the Collaborating Institutions in the Dark Energy Survey. The Collaborating Institutions are Argonne National Laboratory, the University of California at Santa Cruz, the University of Cambridge, Centro de Investigaciones Energeticas, Medioambientales y Tecnologicas-Madrid, the University of Chicago, University College London, the DES-Brazil Consortium, the University of Edinburgh, the Eidgenossische Technische Hochschule (ETH) Zurich, Fermi National Accelerator Laboratory, the University of Illinois at Urbana-Champaign, the Institut de Ciencies de l'Espai (IEEC/CSIC), the Institut de Fisica d'Altes Energies, Lawrence Berkeley National

Laboratory, the Ludwig Maximilians Universitat Munchen and the associated Excellence Cluster Universe, the University of Michigan, NSF's NOIRLab, the University of Nottingham, the Ohio State University, the University of Pennsylvania, the University of Portsmouth, SLAC National Accelerator Laboratory, Stanford University, the University of Sussex, and Texas A&M University.

BASS is a key project of the Telescope Access Program (TAP), which has been funded by the National Astronomical Observatories of China, the Chinese Academy of Sciences (the Strategic Priority Research Program 'The Emergence of Cosmological Structures' Grant # XDB09000000), and the Special Fund for Astronomy from the Ministry of Finance. The BASS is also supported by the External Cooperation Program of Chinese Academy of Sciences (Grant # 114A11KYSB20160057), and Chinese National Natural Science Foundation (Grant # 11433005).

The Legacy Survey team makes use of data products from the Near-Earth Object Wide-field Infrared Survey Explorer (NEOWISE), which is a project of the Jet Propulsion Laboratory/California Institute of Technology. NEOWISE is funded by the National Aeronautics and Space Administration.

The Legacy Surveys imaging of the DESI footprint is supported by the Director, Office of Science, Office of High Energy Physics of the U.S. Department of Energy under Contract No. DE-AC02-05CH1123, by the National Energy Research Scientific Computing Center, a DOE Office of Science User Facility under the same contract; and by the U.S. National Science Foundation, Division of Astronomical Sciences under Contract No. AST-0950945 to NOAO.

The Photometric Redshifts for the Legacy Surveys (PRLS) catalogue used in this paper was produced thanks to funding from the U.S. Department of Energy Office of Science, Office of High Energy Physics via grant DE-SC0007914.

The Siena Galaxy Atlas was made possible by funding support from the U.S. Department of Energy, Office of Science, Office of High Energy Physics under Award Number DE-SC0020086 and from the National Science Foundation under grant AST-1616414.

Funding for the Sloan Digital Sky Survey IV has been provided by the Alfred P. Sloan Foundation, the U.S. Department of Energy Office of Science, and the Participating Institutions. SDSS acknowledges support and resources from the Center for High-Performance Computing at the University of Utah. The SDSS web site is www.sdss4.org.

SDSS is managed by the Astrophysical Research Consortium for the Participating Institutions of the SDSS Collaboration including the Brazilian Participation Group, the Carnegie Institution for Science, Carnegie Mellon University, Center for Astrophysics| Harvard & Smithsonian (CfA), the Chilean Participation Group, the French Participation Group, Instituto de Astrofísica de Canarias, The Johns Hopkins University, Kavli Institute for the Physics and Mathematics of the Universe (IPMU) University of Tokyo, the Korean Participation Group, Lawrence Berkeley National Laboratory, Leibniz Institut für Astrophysik Potsdam (AIP), Max-Planck-Institut für Astronomie (MPIA Heidelberg), Max-Planck-Institut für Astrophysik (MPA Garching), Max-Planck-Institut für Extraterrestrische Physik (MPE), National Astronomical Observatories of China, New Mexico State University, New York University, University of Notre Dame, Observatório Nacional MCTI, The Ohio State University, Pennsylvania State University, Shanghai Astronomical Observatory, United Kingdom Participation Group, Universidad Nacional Autónoma de México, University of Arizona, University of Colorado Boulder, University of Oxford, University of Portsmouth, University of Utah, University of Virginia, University of Washington, University of Wisconsin, Vanderbilt University, and Yale University.

## DATA AVAILABILITY

The CluMPR galaxy cluster catalogue and cluster member galaxy catalogue are currently publicly available at DOI:10.5281/zenodo.10727242. The candidate lensed quasar catalogue is publicly available at DOI:10.5281/zenodo.10727286. Descriptions of each catalogue's columns can be found in Appendix D.

These catalogues are also available to NERSC users at `global/cfs/cdirs/desi/users/mjyb16/CLUMPR_DESI_2024`. The main catalogue can be found in the folder `clean_data`, the extended catalogue including flagged objects can be found in the folder `unclean_data`, catalogues of cluster member galaxies can be found in the folder `members`, and the candidate lensed quasar catalogues can be found in the folder `lensed_qso_candidates`.

Code from this project can be found on GitHub.

## APPENDIX A: GALAXY COUNT AND STELLAR MASS BACKGROUNDS

To determine the backgrounds, we run the CluMPR algorithm as described in 3.1, except that instead of choosing initial points to be the locations of massive galaxies as in 3.1.1, we use random points on the sky. We then calculate the expectation number of galaxies falling within the search cylinder as a function of redshift and take the median across all random points to get a redshift-dependent median galaxy background. A graph of the background galaxy counts (with the median overplotted) is shown in Fig. A1, and a graph of background stellar masses (with median overplotted) is shown in Fig. A2. We perform this process separately for the North (BASS/MzLS) and the South (DECaLS) surveys; the resulting median background galaxy counts and median background stellar masses are shown in Fig. A3 and Fig. A4, respectively.

## APPENDIX B: STELLAR MASS INCOMPLETENESS CORRECTION

The galaxy sample used by our algorithm is magnitude-limited (see 2), which causes a Malmquist bias (Malmquist 1922, 1925). Low-stellar-mass galaxies are thus excluded from our catalogue at a rate which increases with redshift. Thus, in order to obtain a redshift-independent total stellar mass estimate for each cluster, we must correct for the stellar mass incompleteness of our data. To correct for stellar mass incompleteness, we apply a redshift-dependent scaling factor to the galaxy cluster stellar masses. This factor is determined by creating a stellar mass model based on the Schechter function

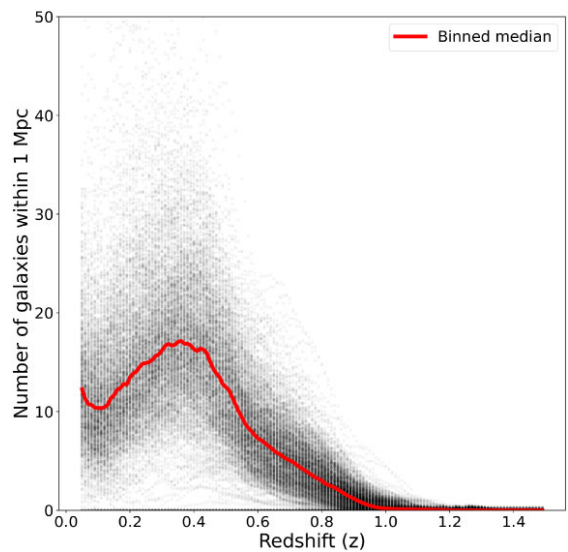

**Figure A1.** Number of galaxies within 1 Mpc at random sky pointings versus photometric redshift. The data in this graph are for the South (DECaLS) portion of the DESI Legacy Survey; the exact same process was applied for the North (BASS/MzLS) portion and yielded similar results. A comparison between the resulting median backgrounds for North and South are shown in Fig. A3.

(Schechter 1976):

$$\Phi(M)dM = \Phi^\star \left(\frac{M}{M_\star}\right)^\alpha \exp\left(\frac{-M}{M_\star}\right) \frac{dM}{M_\star}, \quad (B1)$$

where $M$ is the galaxy stellar mass, $M_\star$ is the characteristic galaxy stellar mass where the power-law form of the function ends, $\Phi(M)$ is the number density of galaxies with a mass $M$, and $\Phi_\star$ is the normalization factor. We fit this model to the high-mass end of the mass distribution histograms of galaxies observed to be in a cluster (each galaxy is weighted based on the probability in the membership catalogues). We are able to use this method since we observe the weighted mass histograms at all redshifts to roughly follow the Schechter function at high masses, though the number count approaches 0 for low-mass galaxies (an example of a stellar mass histogram is shown in Fig. B2).

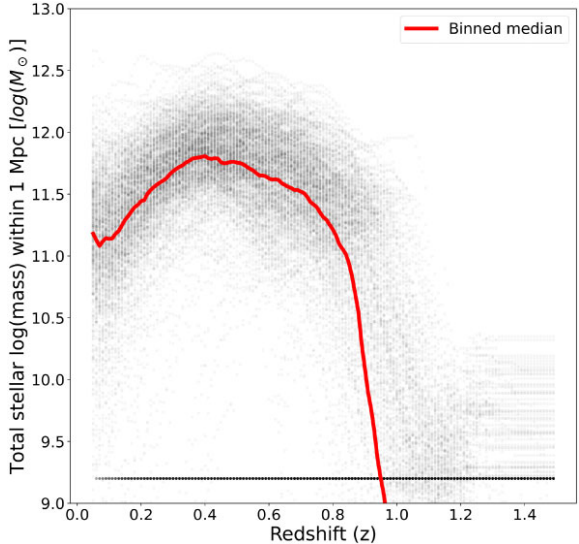

**Figure A2.** 1-Mpc total stellar log(mass) at random sky pointings vs photometric redshift. Pointings which do not contain any galaxies (stellar mass = 0) are plotted at log(mass) = 9.2. For galaxies at $z > 1.2$, stellar masses become unreliable. The data in this graph are for the North (BASS/MzLS) portion of the DESI Legacy Survey; the exact same process was applied for the South (DECaLS) portion and yielded similar results. A comparison between the resulting median backgrounds for North and South are shown in Fig. A4.

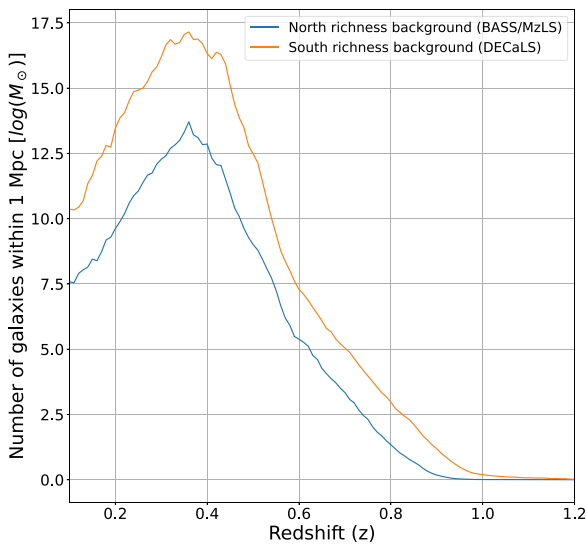

**Figure A3.** Median background galaxy counts within 1 Mpc versus photometric redshift. The differences in backgrounds between North and South are due to differing survey characteristics and different models for redshift for each survey.

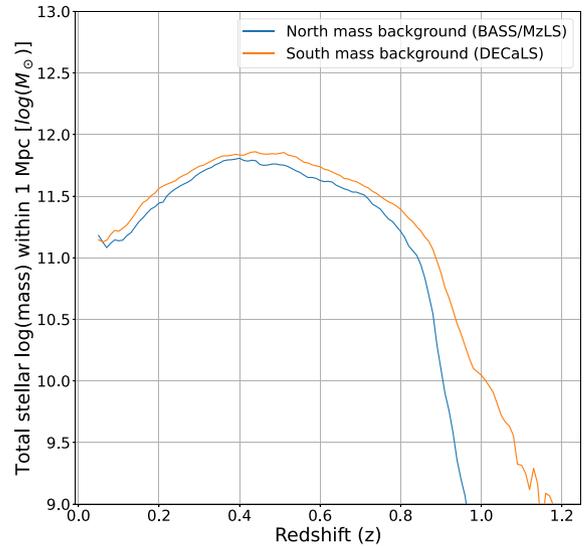

**Figure A4.** Median background 1-Mpc total stellar log(mass) versus photometric redshift. For galaxies at $z > 1.2$, stellar masses become unreliable. The differences in mass backgrounds between North and South are due to differing survey characteristics and different models for stellar mass and redshift for each survey.

As the first step of building our model, we use the Schechter function to approximate the high-mass end of the observed mass distribution of galaxies, though it is somewhat more accurately matched by a double Schechter function (Davidzon et al. 2017;

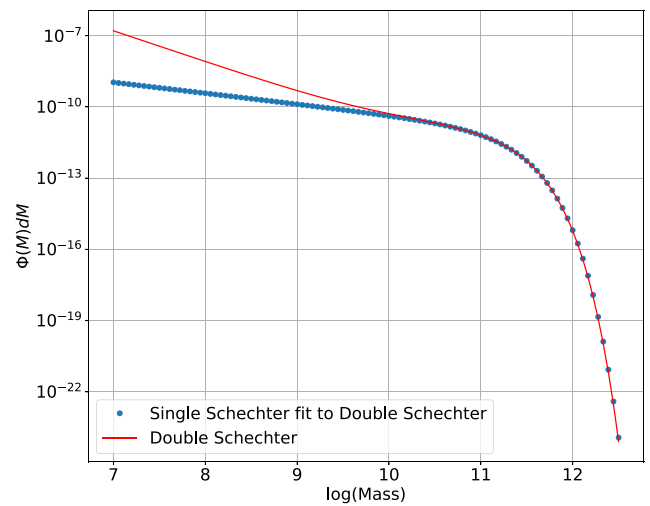

**Figure B1.** Comparison of the double Schechter function with the best-fitting Schechter function as an approximation. The best-fitting Schechter is a very good approximation to the high-mass end of the double Schechter function, down to approximately $\log(M_\odot) = 10$. The parameters for the double Schechter in this example are $\Phi_1^\star = 0.098$, $\alpha_1 = -1.3$, $\Phi_2^\star = 1.58$, $\alpha_2 = -0.39$, $\log(M_\star) = 10.83$. The parameters for the best-fitting Schechter function are $\Phi^\star = 1.663$, $\alpha = -0.459$, $\log(M_\star) = 11.043$. Assuming different parameters for the double Schechter function results in similarly well-fitting Schechter approximations.

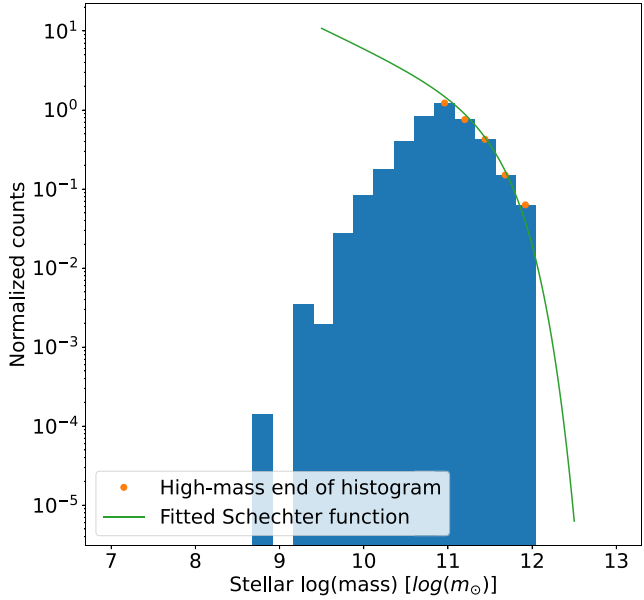

**Figure B2.** An example of fitting a Schechter function to the observed galaxy stellar mass histogram for the redshift bin centred on z = 0.66. The histogram follows a Schechter function above log(mass) = 11, but below this mass the number counts decrease due to incompleteness. The stellar mass correction factor (equation B4) is the ratio between two integrals along the line (Schechter function): the total mass is the integral above log(mass) = 10, and the incomplete mass is the integral above the peak of the histogram (in this case, above log(mass) = 11). This ratio is designed to compensate for the missing galaxy counts for galaxies with masses below log(mass) = 11.

Weaver et al. 2023)

$$\Phi(M)dM = \left(\Phi_1^\star \left(\frac{M}{M_\star}\right)^{\alpha_1} + \Phi_2^\star \left(\frac{M}{M_\star}\right)^{\alpha_2}\right) \exp\left(\frac{-M}{M_\star}\right) \frac{dM}{M_\star}. \quad \text{(B2)}$$

The approximation of a single Schechter function is necessary in order to reduce the number of parameters to be fitted and to make use of the analytic solution to the integral of the single Schechter function. A demonstration of the sufficiency of a single Schechter function to approximate a double Schechter function above log(M) = 10 M$_\odot$ is shown in Fig. B1. By this means, we are able to eliminate what would otherwise be an unconstrained parameter ($\Phi_2$) from our fitting process. We fit for the best $\alpha$ parameter in a single-Schechter model to approximate the observed double Schechter fits for the values shown in Table B1 (Weaver et al. 2023). While the double Schechter function fits we tie to are derived from observations of field galaxies, there is evidence that the observed distribution of galaxy stellar masses does not depend strongly on environment at $z < 1$ (Vulcani et al. 2013). Future work with deeper surveys will need to take into account environmental dependence in luminosity distributions, which becomes more noticeable for $z > 1$ (van der Burg et al. 2020). Unlike $\alpha$, which is fitted to observed stellar mass function shapes and then kept fixed, $M_\star$ and $\Phi(M)$ in our single Schechter model are fitted to the observed stellar mass distribution of galaxies in our clusters. We then evaluate the resulting fits to determine the amount of mass missed due to incompleteness, enabling our cluster stellar mass estimates to be comparable to each other across a range of redshifts.

We fit $M_\star$ and $\Phi^\star$ to the high-mass end of the galaxy mass distribution histogram for several redshift bins (as an example, see Fig. B2). We use 22 redshift bins between $0.025 < z < 1.025$ with bin width $z = 0.05$. The high-mass ends of each histogram are located by fitting a piecewise linear function of the form

Limit log(M$_\odot$) = minimum($a * z + b$, 11.2) (B3)

to the peaks of the histograms, where z is redshift bin. The maximum limit of log(M$_\odot$) = 11.2 is required in order to ensure that the mass of the cluster central galaxy is not excluded from the mass calculation in the main algorithm (see Section 3.2). Note that brightest cluster galaxies (BCGs) have a mass distribution that deviates from the Schechter function (To et al. 2020), so in theory such galaxies ought to be excluded from our fitting process and handled separately. However, in this work we include the cluster central galaxy in the histograms and then fit in log-space to reduce the influence of the highest mass bin, since the central galaxy identified by our algorithm is not necessarily the BCG and hence BCGs cannot be trivially excluded. The bias introduced to the fits by including BCGs should be relatively small for the high-mass clusters in our sample, though it would pose a larger issue for low-mass clusters and groups (Giodini et al. 2009). In future work the BCG can be excluded more precisely for a more accurate fit for low-mass clusters. We find the best-fitting values for equation (B3) to be $a = 1.36202$ and $b = 9.96855$. This model, along with an example of actual histogram peak locations for a small sample of galaxies, is shown in Fig. B3.

The contribution of missing low-mass galaxies to the total stellar mass is determined by taking the ratio between the integral of the single Schechter (equation B1) from log(M$_\odot$) = 10 to infinity and the integral of the single Schechter from the mass limit B3 to infinity. The latter integral represents the incomplete stellar mass calculated by the main algorithm. The ratio between the integrals is thus

factor = $M_{tot}/M_{inc}$, (B4)

where $M_{tot}$ is the total stellar mass above log(M$_\odot$) = 10 and $M_{inc}$ is the incomplete (observed) stellar mass. Multiplying the observed galaxy cluster stellar masses by this redshift-dependent mass correction

**Table B1.** Schechter parameters used to derive stellar mass correction factor.

| | Double Schechter parameters (Weaver et al. 2023) | | | | |
|---|---|---|---|---|---|
| Redshift | log(M$_\odot$) | $\alpha_1$ | $\Phi_1(e5)$ | $\alpha_2$ | $\Phi_2(e5)$ |
| $0.1 < z < 0.5$ | 10.97 | −0.75 | 79.53 | −2.04 | 0.18 |
| $0.5 < z < 0.8$ | 10.90 | −0.50 | 92.06 | −2.14 | 0.09 |
| $0.8 < z < 1.1$ | 10.87 | −0.44 | 102.63 | −2.14 | 0.07 |
| | Best-fitting single Schechter parameters (our model) | | | | |
| Redshift | M$_\odot$ | $\alpha$ | $\Phi$ | | |
| $0.1 < z < 0.5$ | varies with $z$ | −0.7535 | varies with $z$ | – | – |
| $0.5 < z < 0.8$ | varies with $z$ | −0.5015 | varies with $z$ | – | – |
| $0.8 < z < 1.1$ | varies with $z$ | −0.441 | varies with $z$ | – | – |

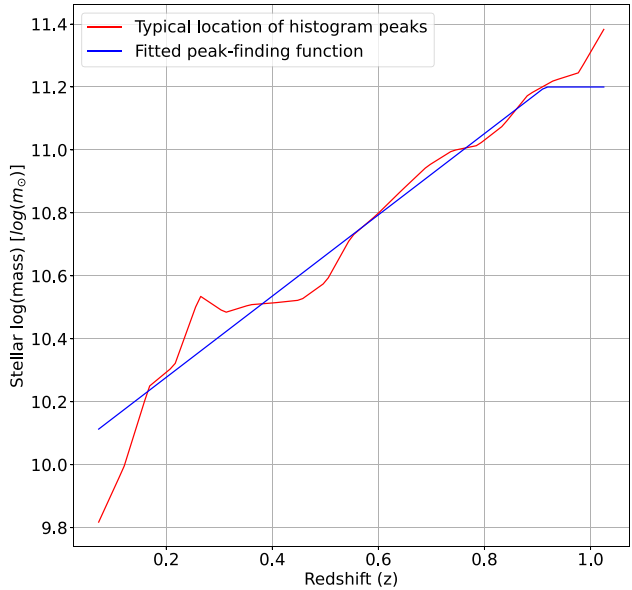

**Figure B3.** The histogram peak-finding function used to determine the log(mass) above which stellar mass histograms should roughly follow a Schechter function. The red (varying-slope) line is the location of histogram peaks for one example sweep; there is good correspondence between the model and the example. In practice, we use many sweeps simultaneously when training the stellar mass correction factor (equation B4), which reduces the random variation in histogram peak location.

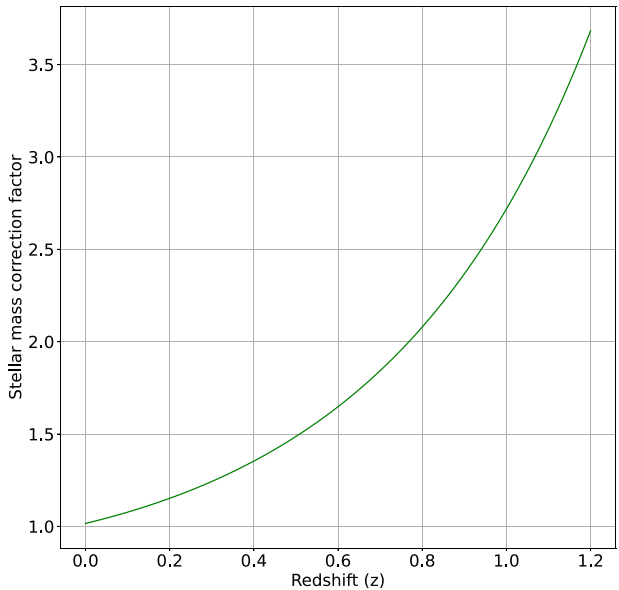

**Figure B4.** Stellar mass correction factor (equation B4) as a function of redshift. Mass incompleteness increases with redshift (higher mass limit), so the correction factor increases correspondingly.

factor allows us to estimate the true stellar masses above $\log(M_\odot) = 10$ within the 1, 0.5, and 0.1 Mpc search radii. We find that a quadratic equation is a good fit for the observed log(stellar mass) correction factor as a function of redshift:

$$\log(\text{factor}(z)) = az^2 + bz + c, \quad \text{(B5)}$$

where $a = 0.19345$, $b = 0.23396$, and $c = 0.00694$. A plot of the stellar mass correction factor as a function of redshift is shown in

Fig. B4. The stellar mass correction factor is provided for all galaxy clusters in our catalogue so that the incomplete stellar mass can be easily retrieved.

## APPENDIX C: GRAVITATIONALLY LENSED QUASAR CANDIDATES

As a supplement to our galaxy cluster catalogue and galaxy cluster member catalogue, we provide a catalogue of candidate gravitationally lensed quasars. Gravitationally lensed quasars are valuable as probes of the Hubble parameter for the expansion of the Universe (Suyu et al. 2017). As data, we used quasars from the DESI Bright and Dark survey target selection (Myers et al. 2023). We compiled the candidate lensed quasar catalogue by counting the number of quasars within the Einstein radius of each galaxy cluster identified by CluMPR. We use two Einstein radii: one corresponds to wide angle lensing by the entire cluster ($M = 10^{15}\,M_\odot$), and the other corresponds to lensing by the core of the cluster ($M = 0.25 * 10^{15}\,M_\odot$). Since there is contamination of QSO targets by stars, particularly in dense star fields, we remove galaxy clusters which have counts within three Einstein radii $>= 6$ times the counts within the target radius. We rate candidates based on how many of the quasars have similar colours in bands g-r, g-z, and r-W1. If at least two quasars are within 1 mag of each other in all three colours, we rate the candidate at Grade C. If at least three quasars are within 1 mag of each other in all three colours, we rate the the candidate at Grade B. If either four quasars or two combinations of three quasars are within 1 mag of each other in all three colours, we rate the candidate as Grade A. Spectroscopic follow-up of the lensed quasar candidates and cluster lens modelling will enable confirmation of true lensed quasars.

### C1 Recovery of known cluster-lensed quasars

To date, six widely separated gravitationally lensed quasars (corresponding to a cluster-scale lens mass) have been discovered (Inada et al. 2003, 2006; Dahle et al. 2013; Shu et al. 2018, 2019; Martinez et al. 2023). Of the six known cluster-lensed quasars, we recover three: SDSS J1004+4112 (Grade A core-lensed), SDSS J1029+2623 (Grade C core-lensed), and COOL J0542-2125 (Grade A core-lensed). We do not recover SDSSJ2222+2745 or SDSS J1326+4806 due to only one of the quasar images being in the DESI target catalogues as a QSO, and we do not recover SDSS J0909+4449 due to the lensing cluster being absent from both the CluMPR and Zou et al. (2021) catalogues (it is more accurately described as a galaxy group).

### C2 New Grade A core-lensed candidates

Of the nine core-lensed candidates ranked Grade A, two are known lensed quasars. We discuss three of the remaining Grade A core-lensed candidates below. The full catalogues of all three grades for both core- and whole-cluster lenses are publicly available as described in the Data Availability section.

#### C2.1 CluMPR J1278885160133433854: a candidate lensed changing-look quasar

The candidate lensed quasars located near the cluster CluMPR J1278885160133433854 are shown in Fig. C1. The galaxy cluster has an SDSS spectroscopic redshift of $z = 0.525$, and the cluster centre galaxy is located at RA = 127.8885, DEC = 43.4339 (J2000). Two of the candidate lensed images, QSO 1 and QSO 2, are confirmed quasars with SDSS redshifts of $z = 1.3056 \pm 0.0002$ and $z = 1.3021 \pm 0.0004$, respectively. The spectra of these quasars are shown in Fig. C2, with zoomed-in views of the C III] and Mg II broad emission lines (Figs C3 and C4, respectively). The alignment between emission lines in the two spectra indicates that QSO 1 and QSO 2 are at the same or nearly the same redshift ($z \approx 1.3$), but the noise in the spectra and the absence of narrow emission lines makes evaluating the accuracy of the assigned SDSS spectroscopic redshifts challenging. There are notable differences between the spectra: QSO 2 appears brighter and has a redder colour (negative slope) compared to QSO 1, and the broad emission lines appear more prominent in QSO 2 than in QSO 1.

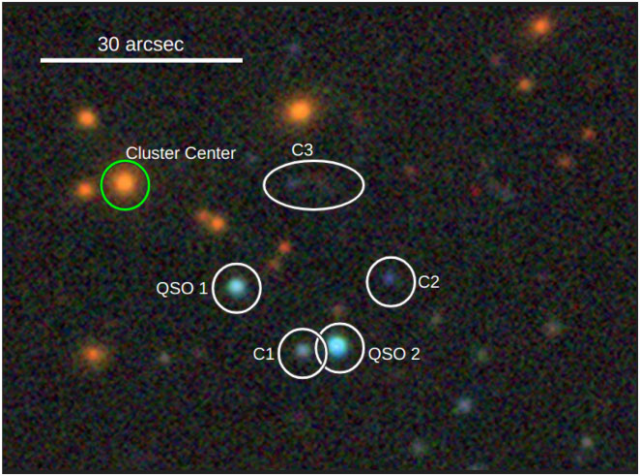

**Figure C1.** Candidate gravitationally lensed quasar system near galaxy cluster CluMPR J1278885160133433854 (cluster centre galaxy in green at far left). QSO 1 and QSO 2 are confirmed quasars with SDSS redshifts of 1.306 and 1.302, respectively. Objects C1 and C2 are QSO targets for DESI. C3 is a very faint object which resembles a lensing arc.

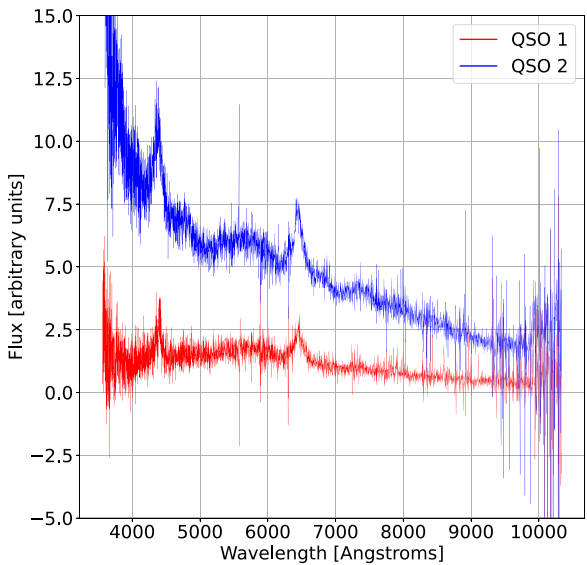

**Figure C2.** SDSS spectra for QSO 1 and QSO 2 in the candidate gravitationally lensed quasar system near galaxy cluster CluMPR J1278885160133433854. The broad emission lines (C III] and Mg II) are redshifted to very similar values near $z = 1.3$, but the slopes of the spectra are very different. This indicates that these are either two different quasars located at the same redshift, or two images of the same highly variable (changing-look) quasar.

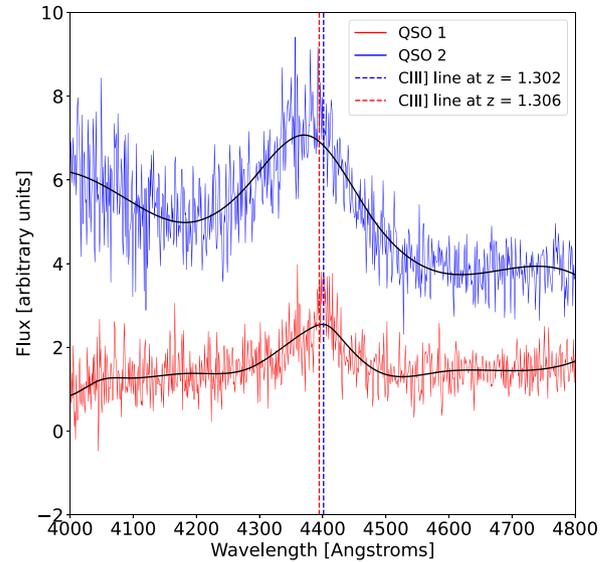

**Figure C3.** Close-up of the C III] broad emission line. QSO 1 has an SDSS redshift estimate of 1.306, while QSO 2 has an SDSS redshift estimate of 1.302. The spectrum for QSO 2 has been shifted down by a constant of 3 flux units in order to improve legibility. The black lines are cubic spline fits to the spectra.

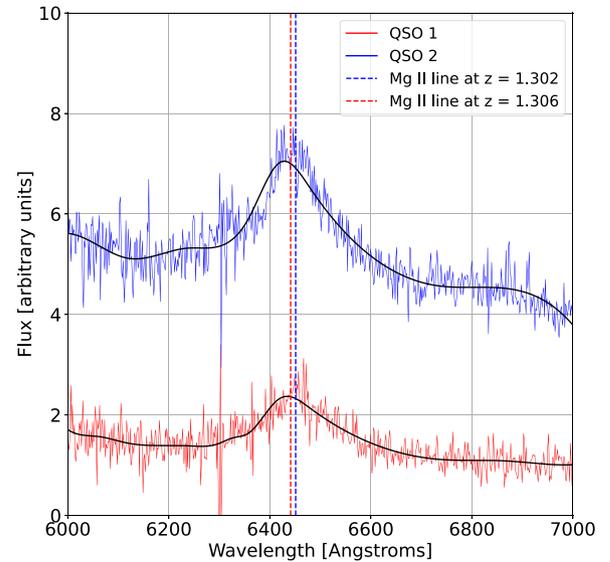

**Figure C4.** Close-up of the Mg II broad emission line. QSO 1 has an SDSS redshift estimate of 1.306, while QSO 2 has an SDSS redshift estimate of 1.302. The black lines are cubic spline fits to the spectra.

We offer two potential explanations for the differences in the spectra: either these are two distinct quasars that happen to be at very similar redshifts, or these are two images of the same highly variable ('changing-look') quasar. Changing-look quasars are quasars whose luminosity and colours (i.e. continuum slopes) change quickly by a large factor over the course of hundreds of days to several years (e.g. Bian et al. 2012; MacLeod et al. 2016, 2019). The luminosity changes are accompanied by the appearance and disappearance of broad emission lines (e.g. LaMassa et al. 2015; Ruan et al. 2016; Runnoe et al. 2016). Changing-look quasars have recently been discovered at redshifts beyond 1 (Guo et al. 2020; Ross et al. 2020). The variability time-scale of changing look quasars is on the same order as the time delays between lensed images for a cluster-scale lens, so it is possible that the difference in spectra between QSO 1 and QSO 2 is due to a gravitational time delay.

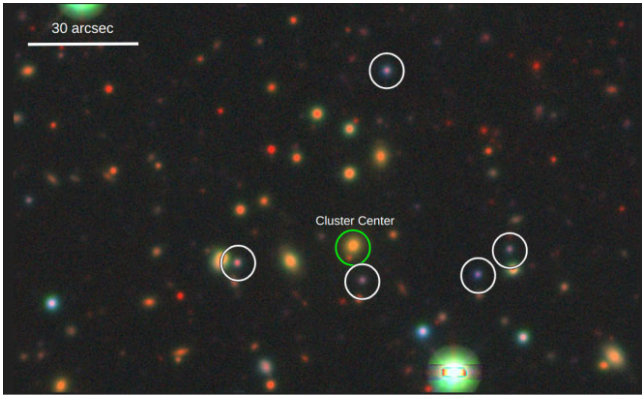

**Figure C5.** Candidate gravitationally lensed quasar system near galaxy cluster CluMPR J2439482900096790366 (cluster centre galaxy in green, second from top in the center). Objects in white circles are QSO targets for DESI, and there is a possibility that they are all images of the same gravitationally lensed quasar.

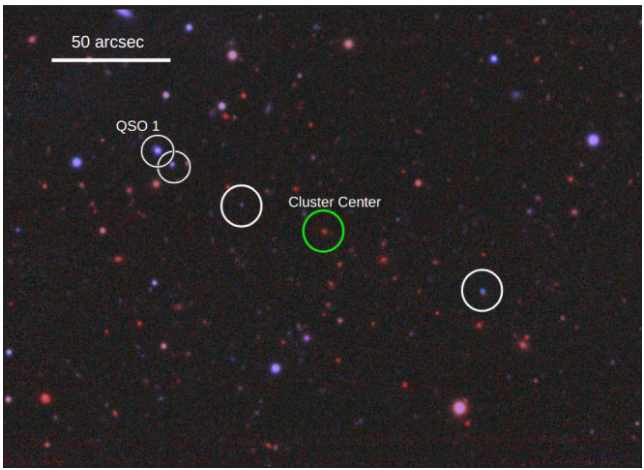

**Figure C6.** Candidate gravitationally lensed quasar system near galaxy cluster CluMPR J2420498300113854205 (cluster centre galaxy in green, second from right). Objects in white circles are QSO targets for DESI, and there is a possibility that some of them are images of the same gravitationally lensed quasar (it is relatively unlikely that they are all the same quasar due to their configuration). QSO 1 is a confirmed quasar with SDSS spectroscopic redshift of 2.205.

To evaluate the changing look quasar interpretation further, we consider the following:

(i) According to the formal uncertainties, the redshifts of the two quasars ($\sim$30–50 km s$^{-1}$) differ by $\approx 8\sigma$. However, the detailed discussion of redshift determination methods used by the SDSS by Lyke et al. (2020, see their section 4.6 and fig. A1) suggests a much larger redshift uncertainty of $\sim$300 km s$^{-1}$. Viewed in this light, the quasar redshifts differ only by 1.5$\sigma$.

(ii) We measured the rest-frame equivalent widths of the broad emission lines and found them to be EW(Mg II) = 34 ± 2 Å and EW(C III]) = 33 ± 3 Å in QSO 1 and EW(Mg II) = 26 ± 1 Å and EW(C III]) = 25 ± 1 Å in QSO 2. These measurements reflect the Baldwin effect (anticorrelation between EW and luminosity; e.g. Green, Forster & Kuraszkiewicz 2001, and references therein) and follow the behaviour seen in extremely variable quasars by Ren et al. (2022).

(iii) A caveat to the above is that, the relative luminosities of the two quasars cannot be readily evaluated as the observed brightness of quasar images depends on the magnification of the lens (Bauer et al. 2011), so the true luminosity of the quasar images cannot be estimated without a well-constrained mass model for the lens.

The above considerations suggest that the lensed changing-look quasar hypothesis remains plausible for QSO 1 and QSO 2. To test this hypothesis further we suggest the following.

(i) *Improving the redshift determination.* – This can be done by obtaining spectra with higher signal-to-noise ratio in the near-IR or sub-mm bands to detect narrow emission lines. Spectra in the wavelength range 7800–8600 Å can catch the narrow [Ne V] $\lambda\lambda 3346, 3426$ and [O II] $\lambda\lambda 3726, 3729$ doublets. The wavelengths of these lines and their relative strengths (the [Ne V]/[O II] ratio) can test the hypothesis that QSO 1 and QSO 2 are lensed images of the same quasar. In the same spirit, spectroscopy in the $J$ band can catch the H$\beta$ line and [O III] $\lambda\lambda 4959, 5007$ doublet and allow a similar test. Alternatively, spectroscopy with ALMA may detect the CO(2-1) line from the quasar host galaxies (rest frequency of 230.5 GHz) that will allow for an exquisite redshift determination.

(ii) *Monitoring for spectroscopic variability.* – If the spectra of QSO 1 and QSO 2 do indeed reflect a change in the state of a single quasar, we should expect that one of the two spectra will change to match the other in the next few years. Therefore, the hypothesis can be tested by repeated spectroscopic observations of the two quasars.

There are two more possible quasar images (C1 and C2) which have been selected for DESI spectroscopy. Finally, there is a dim object resembling a lensing arc. Further spectroscopy and deeper imaging are necessary to determine whether these objects are related to QSO 1 and QSO 2.

### C2.2 CluMPR J2439482900096790366

The candidate lensed quasars located near the cluster CluMPR J2439482900096790366 are shown in Fig. C5. The galaxy cluster's central galaxy is located at the photometric redshift $z = 0.333$ with RA = 243.9483, DEC = 6.7904. There are no publicly available spectra for the candidate quasars at this time, but they are part of the DESI dark time targets. Spectroscopy and deeper imaging are necessary to determine whether this system is a quasar-lensing cluster.

### C2.3 CluMPR J2420498300113854205

The candidate lensed quasars located near the cluster CluMPR J2420498300113854205 are shown in Fig. C6. The galaxy cluster's central galaxy is located at the photometric redshift $z = 0.750$ with RA = 242.0498, DEC = 23.8542. One of the candidate quasars (QSO 1) has SDSS spectroscopy with a spectroscopic redshift of 2.205. There are no publicly available spectra for the other candidate quasars at this time, but they are part of the DESI dark time targets. Spectroscopy and deeper imaging are necessary to determine whether this system is a quasar-lensing cluster.

## APPENDIX D: CATALOGUE DESCRIPTIONS

Below we provide descriptions of the column names and contents for the CluMPR DESI Legacy Survey catalogue (Table D1 and Table D2), associated cluster member catalogue (Table D3), and the candidate gravitationally lensed quasar catalogues (Table D4). The public location of the catalogues can be found in the Data Availability section.

**Table D1.** Main CluMPR galaxy cluster catalogue.

| Columns | Description |
| --- | --- |
| Index | Sequential number in catalogue |
| RA_central | Right ascension of central cluster galaxy (deg) |
| DEC_central | Declination of central cluster galaxy (deg) |
| z_spec | Spectroscopic redshift of central cluster galaxy ($-1$ if unavailable) |
| z_spec_err | Estimated error in spectroscopic redshift of central cluster galaxy ($-1$ if unavailable) |
| z_median_central | Median photometric redshift of central cluster galaxy |
| z_average_no_wt | Mean photometric redshift of all cluster galaxies |
| z_average_prob | Mean photometric redshift of all cluster galaxies, weighted by membership probability |
| z_average_mass_prob | Mean photometric redshift of all cluster galaxies, weighted by membership probability times mass |
| z_std_central | Std. deviation of photometric redshift of central cluster galaxy |
| z_stde_no_wt | Std. error of average photometric redshift of all cluster galaxies |
| z_stde_prob | Std. error of average of photometric redshift of all cluster galaxies, weighted by membership probability |
| z_stde_mass_prob | Std. error of average of photometric redshift of all cluster galaxies, weighted by membership probability times mass |
| gid | Unique identifier of the central cluster galaxy. See Section 3.2 for definition. |
| mass_central | Log(stellar mass) of the central cluster galaxy |
| cluster_mass_onempc | Log(stellar mass) of the cluster within a 1-Mpc physical radius, corrected for incompleteness and galaxy background. 0 indicates that the mass is less than the background. |
| cluster_mass_halfmpc | Log(stellar mass) of the cluster within a 0.5-Mpc physical radius, corrected for incompleteness and galaxy background. 0 indicates that the mass is less than the background. |
| cluster_mass_tenthmpc | Log(stellar mass) of the cluster within a 0.1-Mpc physical radius, corrected for incompleteness and galaxy background. 0 indicates that the mass is less than the background. |
| richness_onempc | Expectation number of observed galaxies in cluster within a 1-Mpc physical radius, corrected for galaxy background |
| richness_halfmpc | Expectation number of observed galaxies in cluster within 0.5-Mpc physical radius, corrected for galaxy background |
| richness_tenthmpc | Expectation number of observed galaxies in cluster within 0.1-Mpc physical radius, corrected for galaxy background |

**Table D2.** Extended CluMPR galaxy cluster catalogue.

| Columns | Description |
| --- | --- |
| Index | Sequential number in catalogue |
| RA_central | Right ascension of central cluster galaxy (deg) |
| DEC_central | Declination of central cluster galaxy (deg) |
| z_spec | Spectroscopic redshift of central cluster galaxy ($-1$ if unavailable) |
| z_spec_err | Estimated error in spectroscopic redshift of central cluster galaxy ($-1$ if unavailable) |
| z_median_central | Median photometric redshift of central cluster galaxy |
| z_average_no_wt | Mean photometric redshift of all cluster galaxies |
| z_average_prob | Mean photometric redshift of all cluster galaxies, weighted by membership probability |
| z_average_mass_prob | Mean photometric redshift of all cluster galaxies, weighted by membership probability times mass |
| z_std_central | Std. deviation of photometric redshift of central cluster galaxy |
| z_stde_no_wt | Std. error of average photometric redshift of all cluster galaxies |
| z_stde_prob | Std. error of average of photometric redshift of all cluster galaxies, weighted by membership probability |
| z_stde_mass_prob | Std. error of average of photometric redshift of all cluster galaxies, weighted by membership probability times mass |
| RELEASE | Integer denoting the camera and filter set used for the central galaxy, which will be unique for a given processing run of the data, from Tractor catalogues |
| BRICKID | A unique Brick ID for the central galaxy, from Tractor catalogues |
| OBJID | catalogue object number within this brick for the central galaxy, from Tractor catalogues |
| MASKBITS | Bitwise mask indicating that the central galaxy touches a pixel in the coadd maskbits maps, from Tractor catalogues |
| gid | Unique identifier of the central cluster galaxy. See Section 3.2 for definition. |
| mass_central | Log(stellar mass) of the central cluster galaxy |
| cluster_mass_onempc | Log(stellar mass) of the cluster within a 1-Mpc physical radius, corrected for incompleteness and galaxy background. 0 indicates that the mass is less than the background. |
| cluster_mass_halfmpc | Log(stellar mass) of the cluster within a 0.5-Mpc physical radius, corrected for incompleteness and galaxy background. 0 indicates that the mass is less than the background. |
| cluster_mass_tenthmpc | Log(stellar mass) of the cluster within a 0.1-Mpc physical radius, corrected for incompleteness and galaxy background. 0 indicates that the mass is less than the background. |

**Table D2** – *continued*

| Columns | Description |
| --- | --- |
| mass_bkgd_onempc | Mass background (which has been subtracted from cluster_mass_onempc) for one-mpc radius |
| mass_bkgd_halfmpc | Mass background (which has been subtracted from cluster_mass_onempc) for one-mpc radius |
| mass_bkgd_tenthmpc | Mass background (which has been subtracted from cluster_mass_onempc) for one-mpc radius |
| correction_factor | Correction factor for mass incompleteness |
| neighbours_onempc | Expectation number of observed galaxies in cluster within a 1-Mpc physical radius, not corrected for galaxy background |
| neighbours_halfmpc | Expectation number of observed galaxies in cluster within a 0.5-Mpc physical radius, not corrected for galaxy background |
| neighbours_tenthmpc | Expectation number of observed galaxies in cluster within a 0.1-Mpc physical radius, not corrected for galaxy background |
| richness_onempc | Expectation number of observed galaxies in cluster within a 1-Mpc physical radius, corrected for galaxy background |
| richness_halfmpc | Expectation number of observed galaxies in cluster within 0.5-Mpc physical radius, corrected for galaxy background |
| richness_tenthmpc | Expectation number of observed galaxies in cluster within 0.1-Mpc physical radius, corrected for galaxy background |
| flag_foreground | 1 indicates the presence of a major foreground galaxy (cluster parameters may be unreliable in such cases). Value 0 should indicate good clusters. |
| edge_mask | 0 indicates that the cluster is located close enough to the edge of the footprint that parameters may be affected. Value 1 should indicate good clusters. |
| flag_footprint | 1 indicates a galaxy cluster which is located in an isolated area of the footprint (these are excluded from the main data set). |

**Table D3.** CluMPR galaxy cluster membership catalogue.

| Columns | Description |
| --- | --- |
| Galaxy | Unique galaxy id, constructed as gid above. See Section 3.2 for definition |
| Cluster | Unique galaxy id of cluster central galaxy |
| Galaxy mass | Stellar mass of galaxy |
| Galaxy redshift | Photometric redshift of galaxy |
| Cluster redshift | Median photometric redshift of cluster central galaxy |
| Galaxy redshift uncertainty | Uncertainty in the photometric redshift of this galaxy |
| Cluster membership probability | Cluster membership probability of this galaxy |

**Table D4.** Candidate lensed quasar catalogues.

| Columns | Description |
| --- | --- |
| lens_RA_central | Right ascension of central cluster galaxy (deg) |
| lens_DEC_central | Declination of central cluster galaxy (deg) |
| lens_z_spec | Spectroscopic redshift of central cluster galaxy ($-1$ if unavailable) |
| lens_z_spec_err | Estimated error in spectroscopic redshift of central cluster galaxy ($-1$ if unavailable) |
| lens_z_median_central | Median photometric redshift of central cluster galaxy |
| cluster_mass_onempc | Log(stellar mass) of the cluster within a 1-Mpc physical radius, corrected for incompleteness and galaxy background |
| lens_gid | Unique identifier of the central cluster galaxy. See Section 3.2 for definition. |
| QSO_RA | Right ascension of each QSO (deg) |
| QSO_DEC | Declination of each QSO (deg) |
| g_r | g-r colour of each QSO |
| g_z | g-z colour of each QSO |
| r_w1 | r-w1 colour of each QSO |

This paper has been typeset from a TEX/LATEX file prepared by the author.